\documentclass[aps,pra,reprint,groupedaddress,nofootinbib]{revtex4-1}

\usepackage{graphicx}
\usepackage{dcolumn}
\usepackage{bm}
\usepackage{amsmath,amssymb,units,mathrsfs}
\usepackage{color}

\def\bra#1#2{\left#1\rule{0ex}{#2ex}\right.}
\begin{document}


\title{Dipole resonances of nonabsorbing dielectric nanospheres in the optical range: Approximate explicit conditions for high- and moderate-refractive-index materials} 

\author{F. L\'opez-Tejeira}
\email[Electronic mail: ]{flt@unizar.es}
\affiliation{Departamento de F\'{\i}sica de la Materia Condensada, Escuela de Ingenier\'{\i}a y Arquitectura, Universidad de Zaragoza, Mar\'{\i}a de Luna 3, E-50018 Zaragoza, Spain}

\date{\today}

\begin{abstract}
In this work, we discuss the way in which electric and magnetic dipole resonances arising in the optical scattering spectrum of non-absorbing dielectric nanospheres can be accurately approximated by means of simple explicit expressions that depend on the sphere's radius, incident wavelength and relative refractive index. We find such expressions to hold not only for high- but also for moderate-refractive-index values, thus complementing the results reported in previous studies.
\end{abstract}

\pacs{}

\maketitle 

\section{Introduction}
\label{sec:intro}

Scattering of light by metallic nanoparticles shows a strongly resonant behavior within the optical region, which permits us to consider them as a sort of optical antennas \cite{Bharadwaj2009}, given their ability to redirect freely propagating light into localized energy, and vice versa. When in the sub-wavelength regime, such resonant nanoparticles may even be used as building blocks for optical meta-materials \cite{Soukoulis2007, Dintinger2012} or meta-objects \cite{Fan2010, Liu2012, Shafiei2013}. Although systems based on metallic nanoparticles have raised the prospect of some very promising applications \cite{Atwater2010, Zhou2016}, they also suffer from two significant limitations when operated within the optical range: they are intrinsically lossy and do not exhibit any intrinsic magnetic response. As a consequence of this, many efforts have been recently devoted to obtain the same functionalities by means of  non-absorbing purely dielectric nanoparticles \cite{Baranov2017, Kruk2017, Yang2017, Tzarouchis2018, Barreda2019, Paniagua-Dominguez2019}, which show magnetic resonances arising from the circulation of light-induced internal displacement currents.

From all possible dielectric nanoparticles, spherical ones are especially well-suited to be used as nanoresonators, given their ease of synthesis by either chemical or physical methods and the fact that Mie theory \cite{Mie1908} explicitly provides the scattering efficiency of a sphere as a function of the incident wavelength, the sphere's radius and its relative refractive index with respect to that of the surrounding medium. Hence, different arrangements have been proposed for purposes of sensing \cite{Garcia-Camara2013, Barreda2015} and directional control of scattered radiation \cite{Gomez-Medina2011, Geffrin2012, Tribelsky2015, Zhang2015, Ullah2018} that are based on the selective excitation of resonances at dielectric nanospheres. In most of these proposals, the obtained scattering response is mainly dominated by dipole resonances, which are those with the lowest energy.

If it were possible to predict the occurrence of a dipole resonance for a triplet of sphere's radius, incident wavelength and refractive index value without the actual evaluation of Mie scattering coefficients, this would undoubtedly result in the easing of nanosphere-based designing. Some previous studies have partially achieved such an objective: Explicit expressions for resonances with any multipolar order and any ordinal number arising in non-absorbing high-refractive index spheres have been presented in a recent paper \cite{Tribelsky2016}. Other authors have obtained similar results for the resonances with the lowest ordinal number of every multipolar order that can be excited in a sphere with a generic real \cite{Roll2000} or complex refractive index \cite{Tzarouchis2016, Tzarouchis2017}. In this work, we discuss the way in which triplets giving rise to electric and magnetic dipole resonances with any ordinal number in non-absorbing spheres can be approximately determined from simple explicit expressions that hold not only for high-refractive-index but also for moderate-refractive-index values, thus complementing those reported in the above-mentioned references.

The paper is structured as follows: In Sec.~\ref{sec:lightscs} we review the basics of light scattering by a non-absorbing dielectric sphere and introduce scattering coefficients and related magnitudes. Section \ref{sec:approxdet} details the way in which dipole resonances arising in the scattering efficiency of high- and moderate-refractive-index spheres can be accurately approximated without the need for full Mie calculation. The validity of these approximations within the optical range for spheres made of Si, $\mathrm{Cu_2O}$ and $\mathrm{TiO_2}$  is discussed in Sec.~\ref{sec:dipolar}. Finally in Sec.~\ref{sec:conclu} we summarize our work.

\section{Scattering of light by a non-absorbing dielectric sphere}
\label{sec:lightscs}

Let us suppose that a uniform, non-magnetic and non-absorbing dielectric sphere with radius $R$ is surrounded by an also non-magnetic and non-absorbing dielectric medium. The dimensionless scattering efficiency for light propagating through the surrounding medium with wavelength $\lambda$ can then be expressed as
\begin{equation}\label{Qsca}
Q_{sca}(x,m)\!=\frac{2}{x^2}\sum_{l=1}^{\infty}\!(2l+1)[|a_l(x,m)|^2+\!|b_l(x,m)|^2],
\end{equation}
where $x=2 \pi R/\lambda>0$ is the size parameter and $m$ the relative refractive index of the sphere with respect to that of the medium \cite{BHBook}. The dependence of $Q_{sca}$ on $m$ (and mostly on $x$) is contained in the scattering coefficients $a_l$ and $b_l$, which represent, respectively, the subsequent electric and magnetic contributions to the multipolar expansion (that is, dipole for $l=1$, quadrupole for $l=2$, octupole for $l=3$, hexadecapole for $l=4$,\ldots) of scattered fields.
\begin{figure}
\includegraphics[width=\columnwidth]{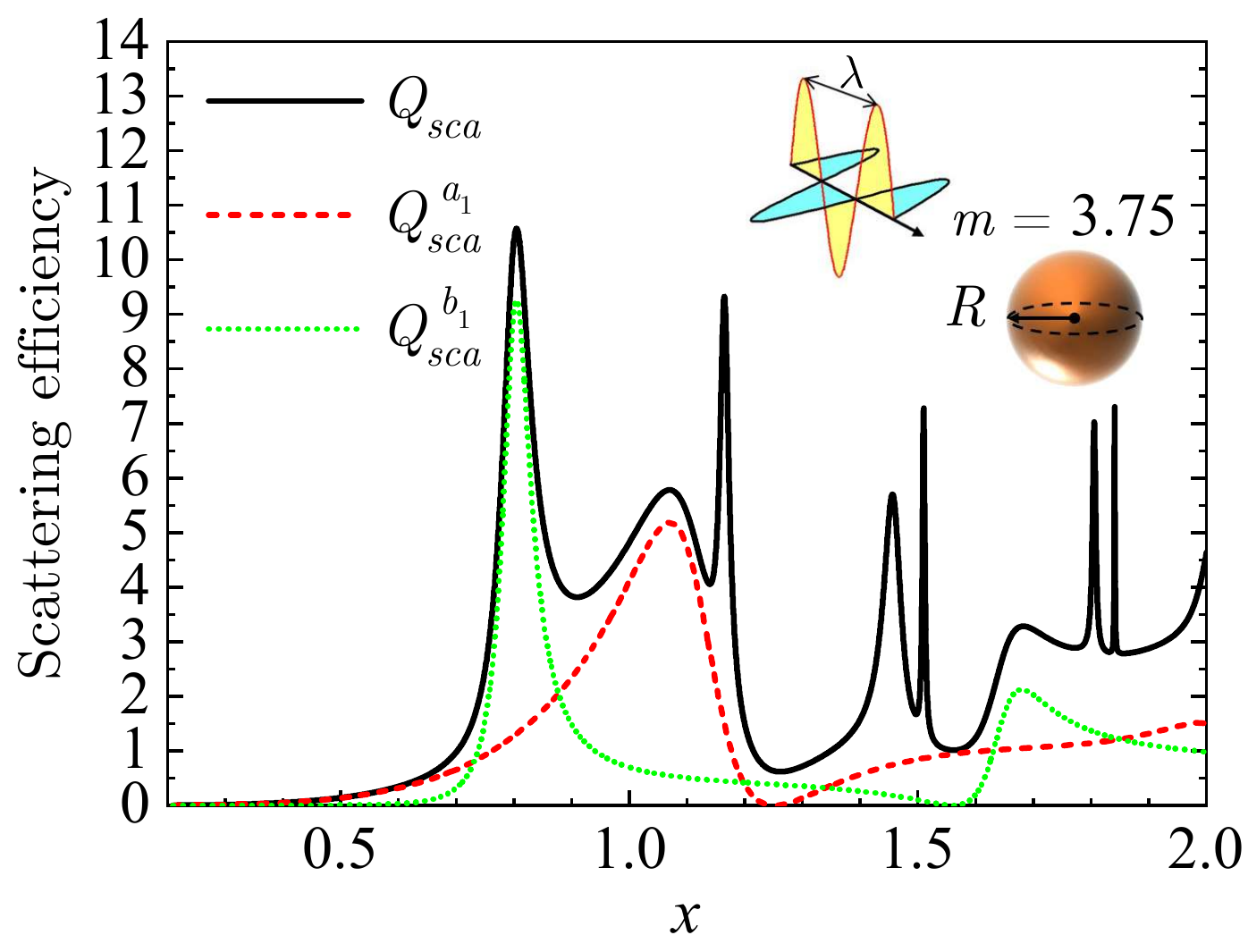}
\caption{\label{fig:Qsca_vs_x} Calculated scattering efficiency $Q_{sca}$ (solid line) as a function of size parameter $x$ for a dielectric sphere with $m=3.75$. Dashed and dotted curves show the values of $Q_{sca}^{a_1}$ and $Q_{sca}^{b_1}$, respectively.}
\end{figure}

Following Refs.~\onlinecite{Chylek1973, Probert-Jones1984}, we find it convenient to write $a_l, b_l$ in the form
\begin{subequations}
\label{scattcoefs}
\begin{equation}
a_l(x,m)=\frac{p_l(x,m)}{p_l(x,m)+iq_l(x,m)},
\end{equation}
\begin{equation}
b_l(x,m)=\frac{r_l(x,m)}{r_l(x,m)+is_l(x,m)},
\end{equation}
\end{subequations}
where
\begin{subequations}
\label{pqrs}
\begin{equation}\label{p}
p_l(x,m)=m \psi_l(mx)\psi_l'(x)-\psi_l(x)\psi_l'(mx),
\end{equation}
\begin{equation}\label{q}
q_l(x,m)=-m \chi_l'(x)\psi_l(mx)+\chi_l(x)\psi_l'(mx),
\end{equation}
\begin{equation}\label{r}
r_l(x,m)=m \psi_l(x)\psi_l'(mx)-\psi_l(mx)\psi_l'(x),
\end{equation}
\begin{equation}\label{s}
s_l(x,m)=-m \chi_l(x)\psi_l'(mx)+\chi_l'(x)\psi_l(mx).
\end{equation}
\end{subequations}

In Eqs.~(\ref{pqrs}), $\psi_l(z)=zj_l(z)$ and $\chi_l(x)=-xy_l(x)$ are the Ricatti-Bessel functions,
which are connected, respectively, to the spherical Bessel functions $j_l$ and $y_l$ \cite{SpecfuncBook}. The prime denotes the derivative with respect to the entire argument of the corresponding function. Please notice that the convenience of writing the scattering coefficients in this fashion rests on the fact that auxiliary functions $p_l, q_l, r_l$ and $s_l$ can only take real values as a direct consequence of $m$'s being real \cite{MishchenkoBook}. This also prevents divergencies in $Q_{sca}$, which shows resonances if either $q_l(x,m)$ or $s_l(x,m)$ vanish. Given $m$ and $l$, there are infinitely many positive values of $x$ that fulfill such conditions, due to the oscillatory nature of the Ricatti-Bessel functions.

As an illustration of this behavior, we present in Fig.~\ref{fig:Qsca_vs_x} the calculated scattering efficiency $Q_{sca}$ (solid line) as a function of size parameter $x$ for a sphere with $m=3.75$, which corresponds to the average value of the range $m=2.5$ to $5$. In order to point out the the origin of different resonances, we also include the specific contribution of dipole terms by means of auxiliary quantities $Q_{sca}^{a_1}=6|a_1|^2/x^2$ (dashed line) and $Q_{sca}^{b_1}=6|b_1|^2/x^2$ (dotted line). As can be seen, dipole contributions dominate the scattering response for $x \lesssim 1.15$, with magnetic and electric resonances located at $x=0.8$ and $1.07$, respectively.

Let us consider that the sphere is made by silicon and surrounded by air. Hence, the incident wavelength for its relative refractive index to be $3.75$ is $720$~nm \cite{Green2008}. This implies that a sphere with radius $R\approx 92$~nm will show a magnetic dipole resonance for that particular wavelength, which, in contrast, will give rise to a mostly electric resonance for $R\approx 123$~nm. If one increases the sphere's radius up to 192~nm (that is $x=1.68$), $Q_{sca}^{b_1}$  will show another peak, although the total scattering efficiency significantly reduces with respect to its value for $x=0.8$. It is then clear that, given two of the three $m$, $\lambda$, and $R$  parameters, dipole resonances can only appear for some specific values of the third one. In the following sections, we will discuss the way in which resonant $(m, \lambda, R)$ triplets can be approximately determined without the actual evaluation of neither $a_1$ nor $b_1$.

\section{Approximate determination of electric and magnetic dipole resonances}
\label{sec:approxdet}

On the assumption that $m$ is kept as a constant, let $\{x_{res}^{a_1,1},x_{res}^{a_1,2},\ldots, x_{res}^{a_1,j},\ldots \}$ be the set of infinitely many positive solutions to
\begin{equation}\label{q1null}
m \chi_1'(x_{res}^{a_1,j})\psi_1(mx_{res}^{a_1,j})=\chi_1(x_{res}^{a_1,j})\psi_1'(mx_{res}^{a_1,j}),
\end{equation}
where $j=1,2,\dots$ is a positive integer number. Hence, an electric dipole resonance appears in $Q_{sca}$ for any pair of values of $R$ and $\lambda$ that meet the condition $R/\lambda=x_{res}^{a_1,j}/2 \pi$, with the caveat that $m$ also depends on $\lambda$. For magnetic dipole resonances, we can then define $\{x_{res}^{b_1,1},x_{res}^{b_1,2},\ldots, x_{res}^{b_1,j},\ldots \}$ as the analog infinite set of positive solutions to
\begin{equation}\label{s1null}
m \chi_1(x_{res}^{b_1,j})\psi_1'(mx_{res}^{b_1,j})=\chi_1'(x_{res}^{b_1,j})\psi_1(mx_{res}^{b_1,j}).
\end{equation}
In order to determine $\{x_{res}^{a_1,j}\}$ and $\{x_{res}^{b_1,j}\}$, let us now take a closer look to the explicit form of the Ricatti-Bessel functions for $l=1$:
\begin{subequations}\label{psi1-chi1p}
\begin{equation}\label{psi1}
\psi_1(mx)=\frac{\sin mx}{mx}-\cos mx,
\end{equation}
\begin{equation}\label{psi1p}
\psi_1'(mx)=-\frac{\sin mx}{(mx)^2}+\frac{\cos mx}{mx}+\sin mx,
\end{equation}
\begin{equation}\label{chi1}
\chi_1(x)=\frac{\cos x}{x}+\sin x,
\end{equation}
\begin{equation}\label{chi1p}
\chi_1'(x)=\cos x-\frac{\cos x}{x^2}-\frac{\sin x}{x}.
\end{equation}
\end{subequations}

It is clearly apparent that Eqs.~(\ref{psi1-chi1p}) can be greatly simplified for some limiting values of $x$ and $m$, thus making it easier to solve Eqs.~(\ref{q1null}) and (\ref{s1null}). In particular, we will consider three different scenarios that are hereafter described in order of increasing complexity.

\subsection{Approximations to $x_{res}^{a_1,j}$ and $x_{res}^{b_1,j}$ for $x \gtrsim 1; m \gg 1$}\label{subsec:1gg2}

 It can be shown from Eqs.~(\ref{psi1-chi1p}) that both $|\chi_1\psi_1'|$ and $|\chi_1'\psi_1|$ are less than one for $x \gtrsim 1$ whatever the value of $m$. For $m \gg 1$, functions $q_1$ and $s_1$ can therefore be approximated as
\begin{subequations}\label{q1s1bigm}
\begin{equation}\label{q1bigm}
q_1(x,m) \approx -m\chi_1'(x)\psi_1(mx),
\end{equation}
\begin{equation}\label{s1bigm}
s_1(x,m) \approx -m\chi_1(x)\psi_1'(mx).
\end{equation}
\end{subequations}
As far as $mx \gg x$, solutions to Eqs.~(\ref{q1null}) and (\ref{s1null}) will then be defined by the following conditions \cite{Tribelsky2016}:
\begin{equation}\label{psi1null}
\psi_1(mx_{res}^{a_1,j})\approx 0,
\end{equation}
\begin{equation}\label{psi1pnull}
\psi_1'(mx_{res}^{b_1,j})\approx 0.
\end{equation}
Let us now assume that $mx$ is large enough to disregard all but purely sinusoidal terms in Eqs.~(\ref{psi1}) and (\ref{psi1p}). Hence, determination of resonances simplifies even more, as size parameters would only have to meet the conditions of Eqs.~(\ref{psi1null}) and (\ref{psi1pnull}) up to the zeroth order in powers of $1/mx$:
\begin{subequations}\label{eq:0thorder}
\begin{equation}
\cos mx_{res (0)}^{a_1,j}=0,
\end{equation}
\begin{equation}
\sin mx_{res (0)}^{b_1,j}=0,
\end{equation}
\end{subequations}
the extra subscript being added in order to avoid any confusion with subsequent results hereafter.

From Eqs.~(\ref{eq:0thorder}), the well-known expressions
\begin{equation}\label{eq:xa1_0}
x_{res (0)}^{a_1,j}(m)=\frac{(j+\frac{\scriptstyle 1}{\scriptstyle 2})\pi}{m},
\end{equation}
\begin{equation}\label{eq:xb1_0}
x_{res(0)}^{b_1,j}(m)=\frac{j\pi}{m}
\end{equation}
are readily obtained. With respect to Eq.~(\ref{eq:xa1_0}), it has to be pointed out that we have set the zeroth-order fundamental resonance to $3 \pi/2m$ (and not to $\pi/2m$) in order to be consistent with $x \gtrsim 1$. Nevertheless, we would rather preserve the full version of Eqs.~(\ref{psi1null}) and (\ref{psi1pnull}) and find their solutions by expanding in a Taylor series about $x_{res (0)}^{a_1,j}$ and $x_{res (0)}^{b_1,j}$ [Cf. Eqs.~(8.13) and (8.14) in Ref.~\onlinecite{Tribelsky2016}]:
\begin{equation}\label{xa1Trib}
x_{res(1)}^{a_1,j}(m)=x_{res (0)}^{a_1,j}(m)\bra{[}{3}\!\!1-\frac{1}{(j+\frac{\scriptstyle 1}{\scriptstyle 2})^2 \pi^2-1}\!\bra{]}{3},
\end{equation}
\begin{equation}\label{xb1Trib}
x_{res(1)}^{b_1,j}(m)=x_{res(0)}^{b_1,j}(m)\bra{[}{3}\!\!1-\frac{1}{j^2 \pi^2-2}\!\bra{]}{3}.
\end{equation}
As far as the obtained expressions are nothing other than the sequential zeros of $j_1(mx)$ and $mx j_0(mx)-j_1(mx)$, their numerical precision can be extended on demand by means of standard techniques \cite{SpecfuncBook}.
\begin{figure}
\includegraphics[width=\columnwidth]{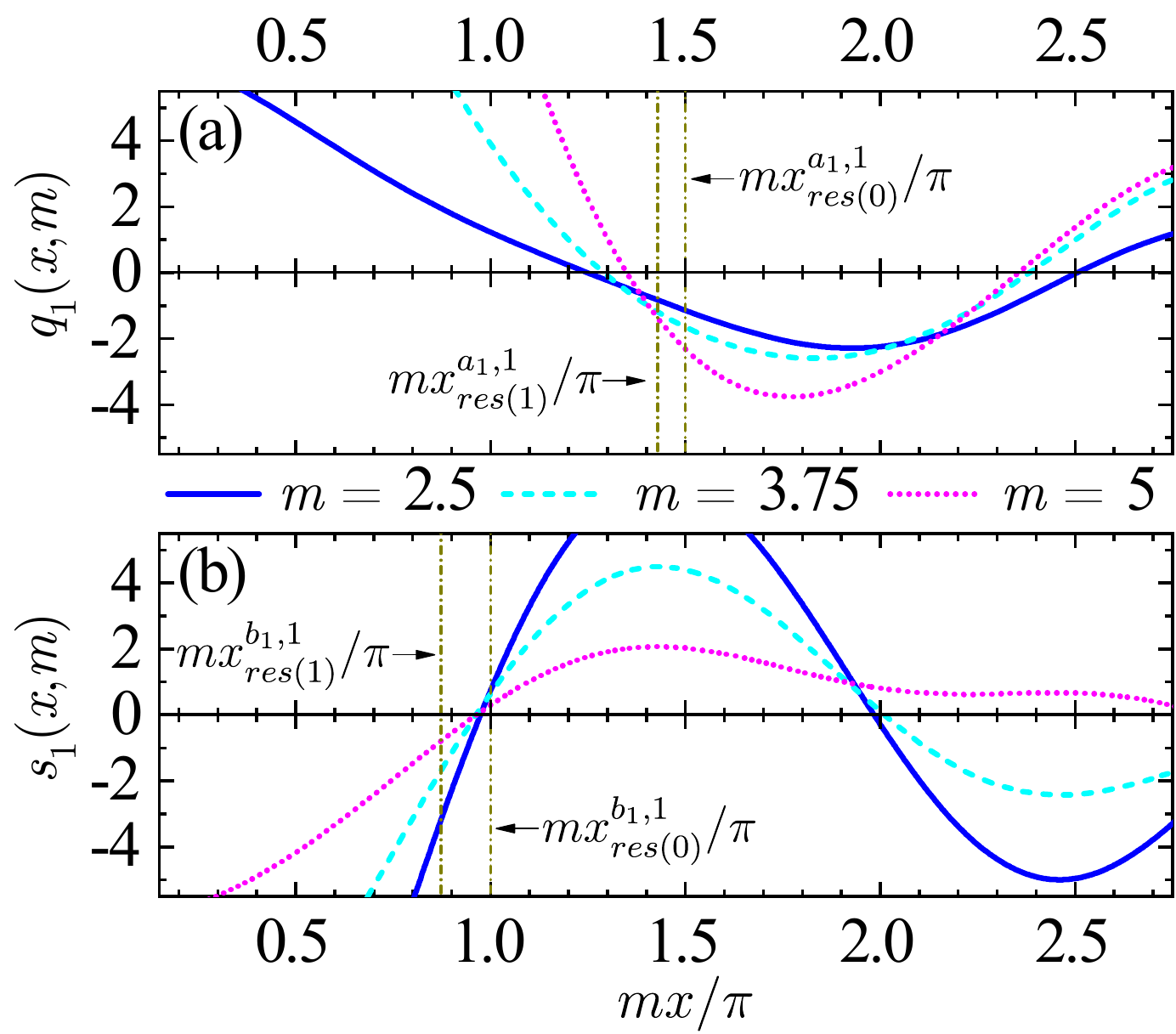}
\caption{\label{fig:q1s1} Calculated $q_1(x,m)$ (a) and $s_1(x,m)$ (b) as a function of $mx/\pi$. Solid, dashed and dotted curves show the values for $m=2.5$, $3.75$, and $5$, respectively. Vertical dashed-dotted lines mark the values of $x_{res(0)}^{a_1,1}$, $x_{res(1)}^{a_1,1}$, $x_{res(0)}^{b_1,1}$ and $x_{res(1)}^{b_1,1}$ in units of $\pi/m$.}
\end{figure}

\subsection{Improved approximations to $x_{res}^{a_1,1}$ and $x_{res}^{b_1,1}$}\label{subsec:a11b11}

According to Eq.~(\ref{xa1Trib}), we expect the size parameter for fundamental electric dipole resonance to be approximately equal to $1.43$ in units of $\pi/m$. However, as can be seen in Fig.~\ref{fig:q1s1}(a), that value actually defines some upper bound that is not reached even for $m=5$. For the fundamental magnetic dipole resonance in Fig.~\ref{fig:q1s1}(b), $x_{res(0)}^{b_1,1}$ seems to provide a better approximation than $x_{res(1)}^{b_1,1}$, although neither of them completely captures the dependence of $x_{res}^{b_1,1}$ on $m$. It is then clear that assumptions made in Sec.~\ref{subsec:1gg2} are too restrictive to provide accurate results for the position of fundamental dipole resonances when $m$ lies between $2.5$ and $5$. We will therefore attempt a different approach.

\begin{figure}
\includegraphics[width=\columnwidth]{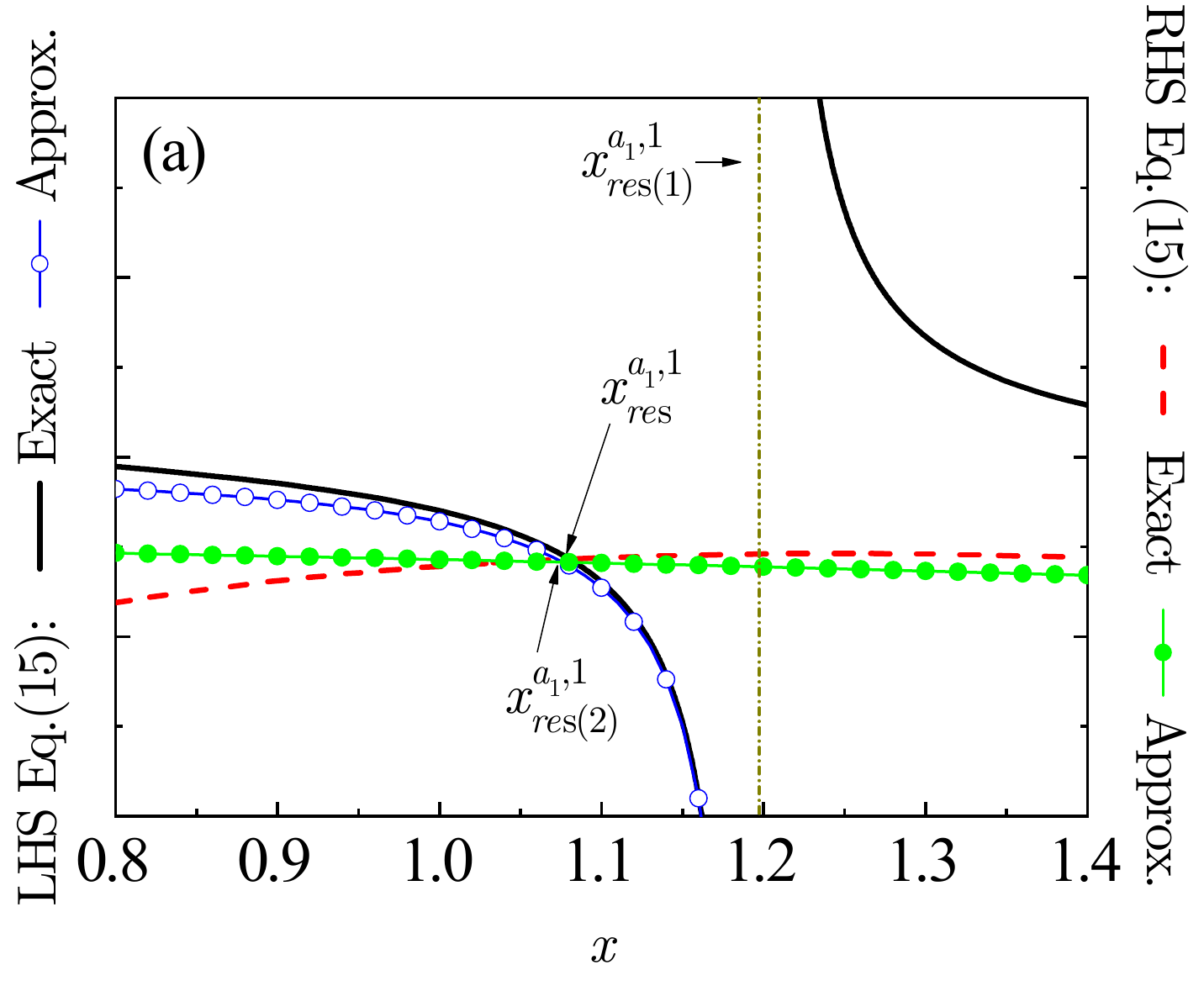}
\includegraphics[width=\columnwidth]{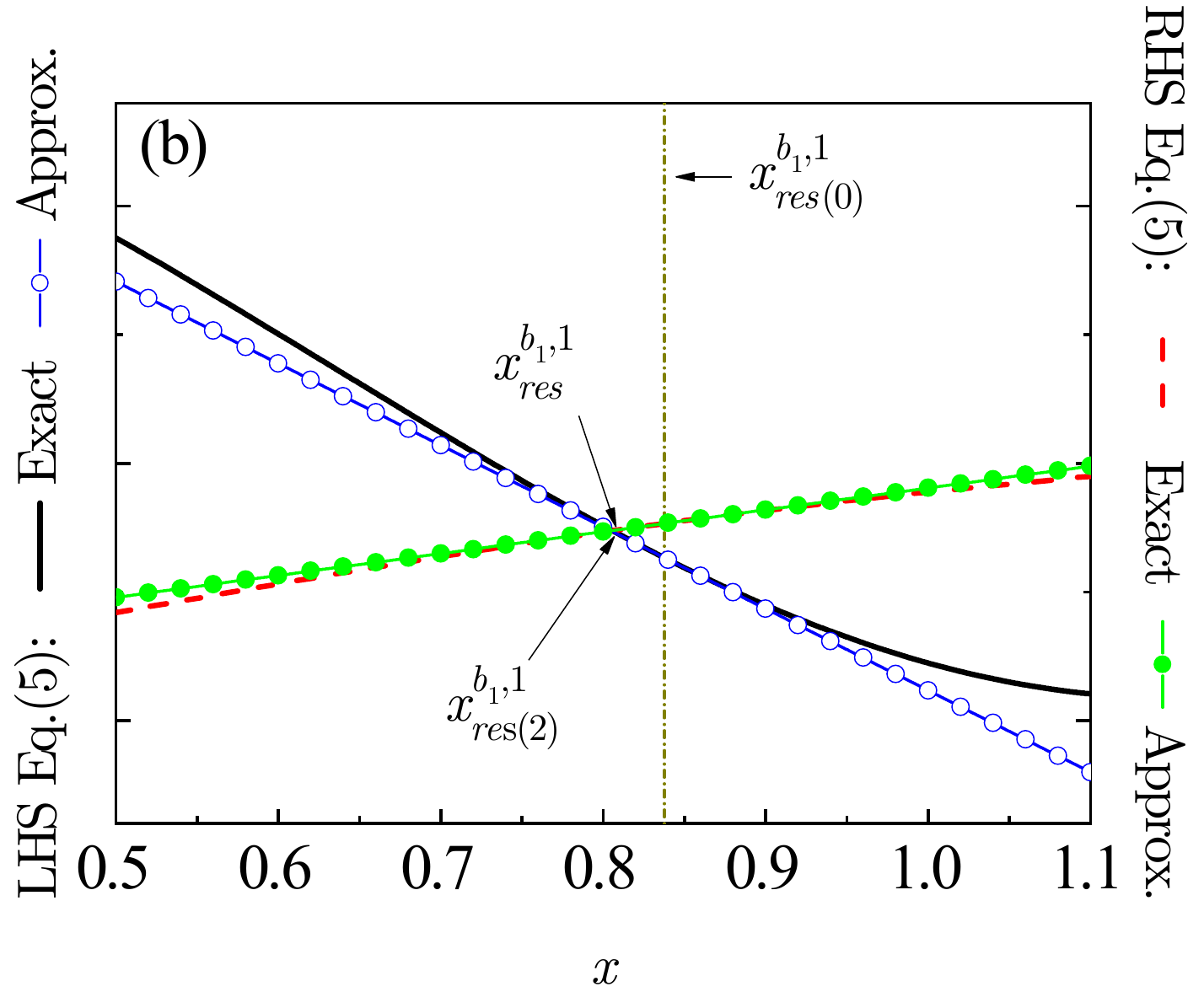}
\caption{\label{fig:graphsols} Graphical determination of the fundamental electric (a) and magnetic (b) dipole resonances for $m=3.75$. Solid and dashed curves in (a) show the exact values for left- and right-hand sides of Eq.~(\ref{eq:q1nullbis}), respectively. Open ($\circ$) and solid ($\bullet$) symbols denote their proposed approximations (see text). Same conventions apply for Eq.~(\ref{s1null}) in (b). Vertical dashed-dotted lines mark the positions of $x_{res(1)}^{a_1,1}$ and $x_{res(0)}^{b_1,1}$.}
\end{figure}

Let us keep all the terms in Eqs~(\ref{q1null}) and slightly recast them so that the two kinds of Ricatti-Bessel functions are separated:
\begin{equation}\label{eq:q1nullbis}
\frac{1}{m}\frac{\psi_1'(mx_{res}^{a_1,1})}{\psi_1(mx_{res}^{a_1,1})}=\frac{\chi_1'(x_{res}^{a_1,1})}{\chi_1(x_{res}^{a_1,1})}.
\end{equation}
In Fig.~\ref{fig:graphsols}(a) we present the graphical solution to Eq.~(\ref{eq:q1nullbis}) for $m=3.75$. As can be seen, the intersection of $\psi_1'/m\psi_1$ (solid line) and $\chi_1'/\chi_1$ (dashed line) takes place for some $x_{res}^{a_1,1}$ that is located in the vicinity of $x_{res(1)}^{a_1,1}$, which is the first positive zero of $\psi_1(x,m)$. Therefore, the solution of Eq.~(\ref{eq:q1nullbis}) is close to an infinite discontinuity (pole) of $\psi_1'/\psi_1$. We can then replace such a function by its [0/1] Pad\'e approximant \cite{PadeBook} at $x=x_{res(1)}^{a_1,1}(m)$, that is
\begin{equation}\label{eq:padepsi1}
\frac{1}{m}\frac{\psi_1'(mx)}{\psi_1(mx)}\approx \frac{1}{m^2 \left(x-x_{res(1)}^{a_1,1}(m)\right)}.
\end{equation}
With respect to the right-hand side of Eq.~(\ref{eq:q1nullbis}), we find it convenient to make use of a [0/1] economized rational approximation (ERA) \cite{NRBook} to $\chi_1'(x)/\chi_1(x)$ over the interval $x_{res(1)}^{a_1,1}(5)\approx 0.9$ to $x_{res(1)}^{a_1,1}(2.5)\approx 1.8$. We therefore obtain an error distribution that is more uniform than that of the corresponding Pad\'e approximant about the midpoint:
\begin{equation}\label{eq:chi1ratio}
\chi_1'(x)/\chi_1(x) \approx -\frac{12}{(28-7x)}.
\end{equation}
From the right-hand sides of Eqs.~(\ref{eq:padepsi1}) and (\ref{eq:chi1ratio}) (which are represented in Fig.~\ref{fig:graphsols}(a) by open and solid symbols, respectively), we finally arrive to an improved explicit expression for the position of the fundamental electric dipole resonance:
\begin{equation}\label{eq:xa112}
x_{res(2)}^{a_1,1}(m)=\frac{x_{res(1)}^{a_1,1}(m)-\frac{{\displaystyle 7}}{{\displaystyle 3m^2}}}{1-\frac{{\displaystyle 7}}{{\displaystyle 12m^2}}}.
\end{equation}

For the determination of the fundamental magnetic dipole resonance, we will keep Eq.~(\ref{s1null}) in its original form. As previously shown in Fig.~\ref{fig:q1s1}(b), $x_{res}^{b_1,1}$ is very close to $x_{res(0)}^{b_1,1}$ (in fact, $x_{res}^{b_1,1}$ is exactly equal to $\pi/2$ for $m=2$). We can therefore approximate each side of Eq.~(\ref{s1null}) by its corresponding linear Taylor expansion about $x=\pi/m$:
\begin{subequations}
\begin{equation}\label{tayloreq5L}
 m \psi_1'(mx)\chi_1(x) \approx L_0(m)+L_1(m)\left(x-\frac{\pi }{m}\right)
\end{equation}
\begin{equation}\label{tayloreq5R}
\chi_1'(x)\psi_1(mx) \approx D_0(m)+D_1(m)\left(x-\frac{\pi }{m}\right)
\end{equation}
\end{subequations}
where
\begin{subequations}
\begin{equation}
L_0(m)=-\frac{m^2 }{\pi ^2}\cos \frac{\pi }{m}-\frac{m }{\pi }\sin \frac{\pi }{m},
\end{equation}
\begin{eqnarray}
L_1(m) &=& \left(\frac{m^3 }{\pi ^3}(3-\pi^2)-\frac{m}{\pi}\right)\!\cos \frac{\pi }{m}+{}\nonumber\\
&+&\left(\frac{m^2 }{\pi^2}(3-\pi^2) \right)\!\sin \frac{\pi }{m},
\end{eqnarray}
\begin{equation}
D_0(m)=\left(1-\frac{ m^2}{\pi ^2}\right)\cos \frac{\pi}{m}-\frac{m }{\pi}\sin \frac{\pi }{m},
\end{equation}
\begin{eqnarray}
D_1(m)& =&\left(\frac{3 m^3 }{\pi ^3}-\frac{2 m }{\pi }\right)\cos \frac{\pi }{m}+{} \nonumber\\ & + &\left(\frac{3 m^2}{\pi ^2}-1\right)\sin \frac{\pi }{m}.
\end{eqnarray}
\end{subequations}
The intersection of the right-hand sides of Eqs.~(\ref{tayloreq5L}) and (\ref{tayloreq5R}) provides an excellent approximation to $x_{res}^{b_1,1}$, as can be seen in Fig.~\ref{fig:graphsols}b for $m=3.75$ (open and solid symbols). By solving this linear form of  Eq.~(\ref{s1null}), the position of the resonance can then be expressed as
\begin{equation}\label{eq:xb112}
x_{res(2)}^{b_1,1}(m)=x_{res(0)}^{b_1,1}(m)+\Delta x_{res}^{b_1,1} (m),
\end{equation}
where
\begin{equation}\label{deltaxb10}
\Delta x_{res}^{b_1,1}(m)= -\frac{\pi}{(m^2-1)(m+\pi \tan \frac{\pi}{m})}.
\end{equation}

\begin{figure}
\includegraphics[width=\columnwidth]{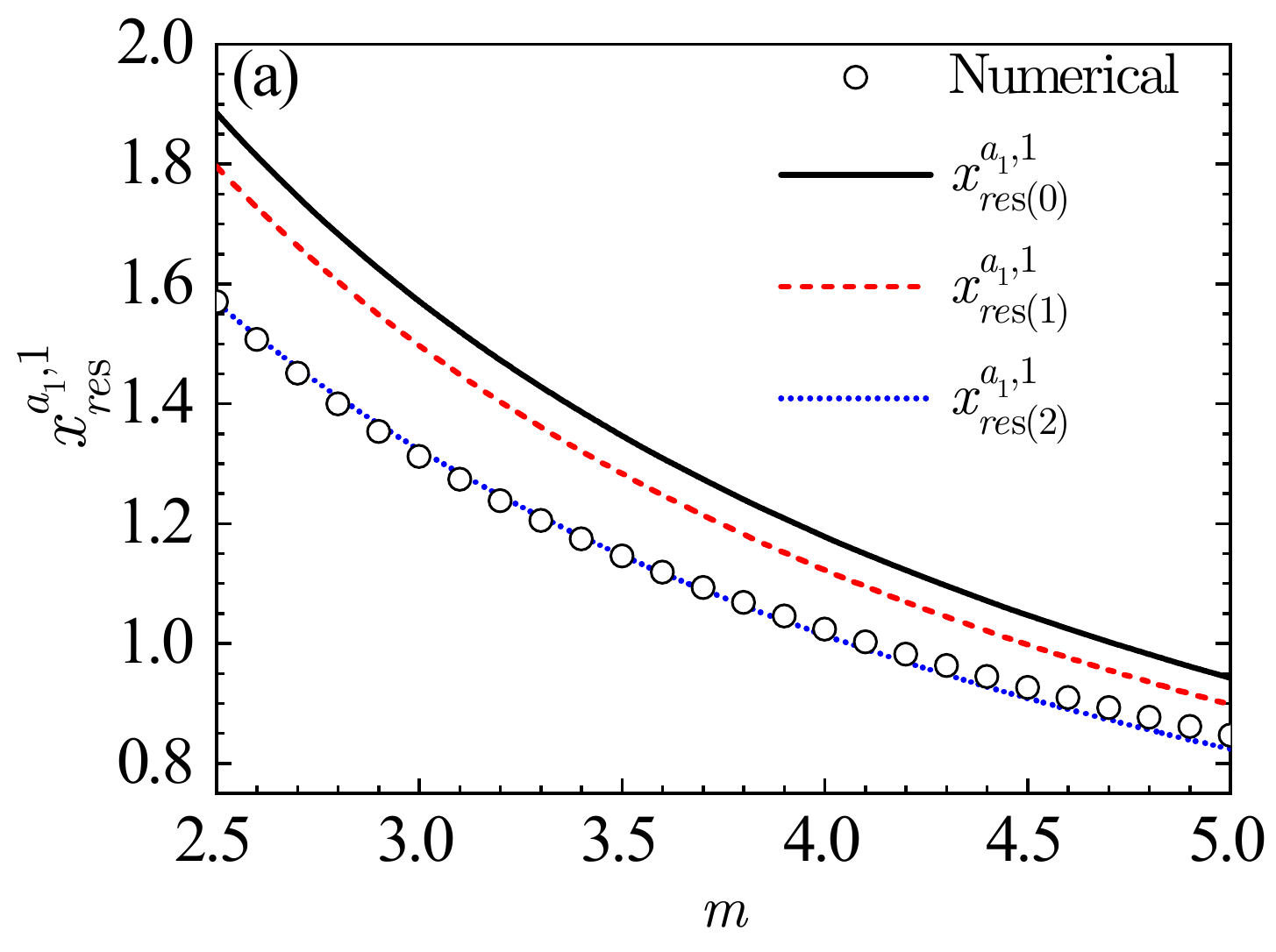}
\includegraphics[width=\columnwidth]{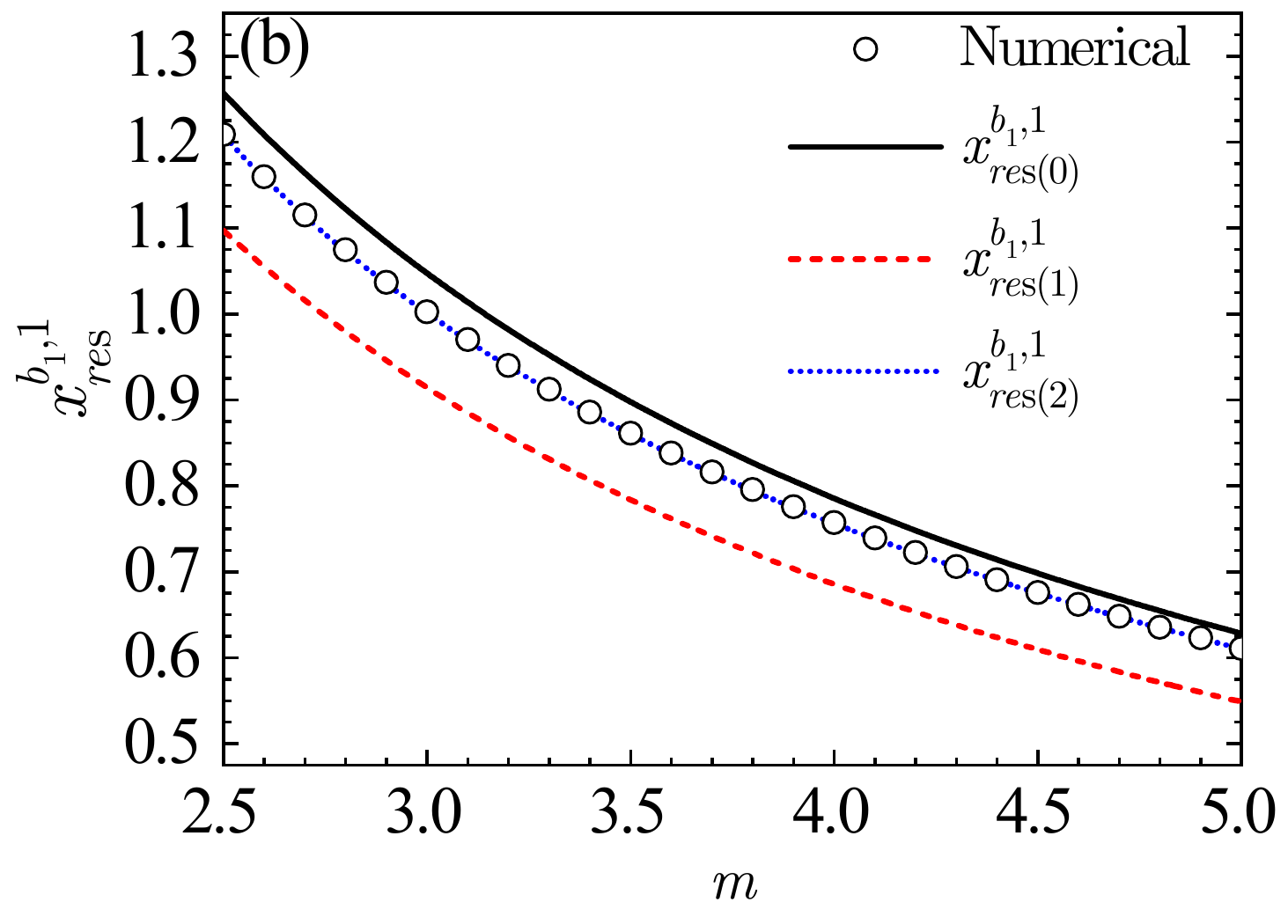}
\caption{\label{fig:xresvsm} Calculated size parameter as a function of $m$ for the fundamental electric (a) and magnetic (b) dipole resonances. Open symbols ($\circ$) in (a) correspond to the numerical solution to Eq.~(\ref{q1null}), whereas solid, dashed, and dotted curves show the values of the proposed approximations to $x_{res}^{a_1,1}$ with subscripts $(0)$, $(1)$ and $(2)$ (see text). Same conventions apply for Eq.~(\ref{s1null}) and $x_{res}^{b_1,1}$ in panel (b).}
\end{figure}

Figure~\ref{fig:xresvsm} shows the calculated values of size parameter as a function of $m$ for the fundamental electric and magnetic dipole resonances of a non-absorbing dielectric sphere with its relative refractive index between $2.5$ and $5$. Open symbols ($\circ$) denote the numerical solutions to Eqs.~(\ref{q1null}) and (\ref{s1null}), whereas solid, dashed, and dotted lines show the values of the proposed approximations to $x_{res}^{a_1,1}$ and $x_{res}^{b_1,1}$ with subscripts $(0)$, $(1)$ and $(2)$ respectively. As can be seen in Fig.~\ref{fig:xresvsm}(a), the size parameter corresponding to the fundamental electric dipole resonance ranges between $1.57$ for $m=2.5$ and $0.85$ for $m=5$. These values are systematically overestimated by $x_{res(0)}^{a_1,1}$, which bears a percentage error that increases from $\%_{error}=+11$ for $m=5$ to $\%_{error}=+20$ for $m=2.5$. As regards $x_{res(1)}^{a_1,1}$, one may observe that it shows an acceptable $\%_{error}=+6$ for $m=5$ and then becomes less reliable as $m$ decreases ($\%_{error}=+14$ for $m=2.5$). On the other hand, proposed $x_{res(2)}^{a_1,1}$ keeps the absolute value of percentage error below $3$ for the entire range of refractive index values, thus providing a significant improvement to the approximate determination of $x_{res}^{a_1,1}$.

With respect to the fundamental magnetic dipole resonance, we have already mentioned that $x_{res(0)}^{b_1,1}$ provides an estimate for $x_{res}^{b_1,1}$ that is exact for $m=2$. As shown in Fig.~\ref{fig:xresvsm}b, $x_{res(0)}^{b_1,1}$ slightly overestimates the size parameter for $m$ above $2$, although its percentage error does not exceed $6$ for any considered value of $m$. Unlike $x_{res(1)}^{a_1,1}$, $x_{res(1)}^{b_1,1}$ does not improve the approximation to the resonance. Conversely, it consistently underestimates the resonant size parameter throughout the interval with a percentage error that ranges between $\%_{error}=-9$ for $m=2.5$ and $\%_{error}=-10$ for $m=5$. Fortunately, the $x_{res(2)}^{b_1,1}$ approximation is in turn found to be $99\%$ accurate for the range between $m=2.5$ and $m=5$, which shows the convenience of this approach.
\begin{figure*}
\includegraphics[width=2\columnwidth]{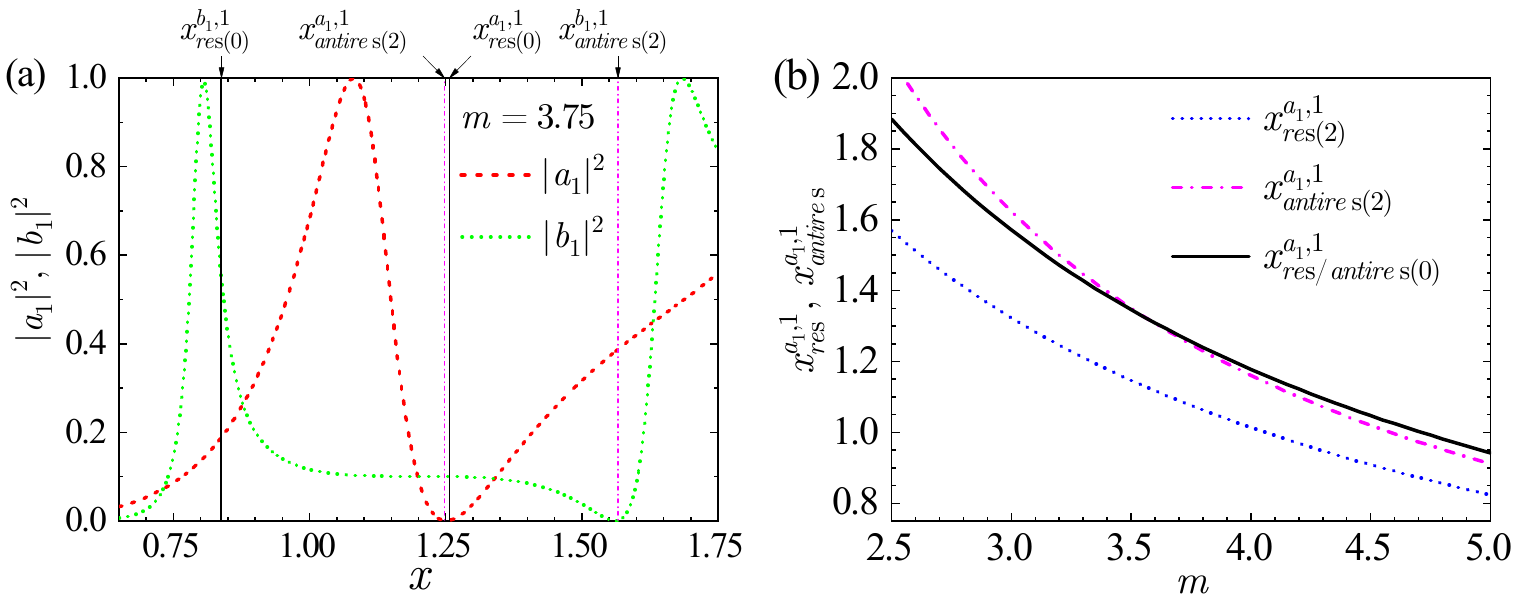}
\caption{\label{fig:resantires} (a) Calculated values of $|a_1|^2$ (dashed) and $|b_1|^2$ (dotted) as a function of size parameter $x$ for a dielectric sphere with $m=3.75$. Vertical solid lines show the approximated values of fundamental dipole resonances with subscript $(0)$, whereas dashed-dotted ones mark the positions of approximated antiresonances with subscript $(2)$. (b) Calculated size parameters as a function of $m$ for the approximations with subscript $(2)$ to the fundamental electric dipole resonance (dotted) and antiresonance (dashed-dotted). Solid curve shows the values of proposed approximations with subscript $(0)$.}
\end{figure*}

In order to better understand the very different reliability of approximations with subscript $(0)$ for the fundamental electric and magnetic dipole resonances, let us now consider an issue that seems at first sight unrelated to the subject, namely the determination of fundamental dipole antiresonances. If either $|a_1|^2$ or $|b_1|^2$ vanish for $m>1$, dipole contributions to scattering are then suppressed, thus producing a noticeable deep in $Q_{sca}$ (see, e.g., Fig.~\ref{fig:Qsca_vs_x}) unless other multipolar orders be dominant. Dipole resonances and their antagonists are, by the way, very close to each other so that plots of $|a_1|^2$ and $|b_1|^2$ as a function of either $x$ or $m$ exhibit a Fano-type line shape \cite{Tribelsky2012, Arruda2015, Tribelsky2016}. Aside from their fundamental interest, dipole antiresonances may also be relevant by themselves for the designing of dielectric nanoresonators \cite{Miroshnichenko2015}.

According to Eq.~(\ref{scattcoefs}a), an electric dipole antiresonance is expected to happen whenever $p_1(x,m)$ is equal to zero. By following exactly the same procedure as in Sec.~\ref{subsec:1gg2}, we obtain that approximations with subscripts $(0)$ and $(1)$ for the fundamental electric dipole antiresonance do coincide with those of $x_{res}^{a_1,1}$:
\begin{equation}\label{eq:xantia10}
x_{antires(0)}^{a_1,1}(m)=x_{res(0)}^{a_1,1}(m)=\frac{3 \pi}{2m},
\end{equation}
\begin{equation}\label{eq:xantia11}
x_{antires(1)}^{a_1,1}(m)=x_{res(1)}^{a_1,1}(m)=\frac{3 \pi}{2m}\frac{(9 \pi^2-8)}{(9 \pi^2-4)}.
\end{equation}
It is not but up to approximation with subscript $(2)$ (that is, [0/1] Pad\'e for $\psi'_1(mx)/m\psi_1(mx)$ and [0/1] ERA for $\psi'_1(x)/\psi_1(x)$ ) that size parameters of resonance and antiresonance depart from each other:
\begin{equation}\label{eq:xantia12}
x_{antires(2)}^{a_1,1}(m)=\frac{x_{antires(1)}^{a_1,1}(m)-\frac{{\displaystyle 5}}{{\displaystyle 7m^2}}}{1-\frac{{\displaystyle 8}}{{\displaystyle 7m^2}}}.
\end{equation}

In contrast, approximation with subscript $(0)$ for the fundamental magnetic dipole antiresonance leads us to
\begin{equation}\label{eq:xantib110}
x_{antires(0)}^{b_1,1}(m)=\frac{2 \pi}{m}
\end{equation}
and not to $\pi /m$. The subsequent linear expansion of
\begin{equation}
r_1(x,m)=0
\end{equation}
about $x_{antires(0)}^{b_1,1}$ allows one to obtain
\begin{equation}\label{eq:xantib112}
x_{antires(2)}^{b_1,1}(m)=x_{antires(0)}^{b_1,1}(m)+\Delta x_{antires}^{b_1,1}(m),
\end{equation}
with
\begin{equation}\label{delta_antixb112}
\Delta x_{antires}^{b_1,1}(m) \approx \frac{58 (99-46 m)}{5 \left(1100 m^2-2155 m+425\right)}.
\end{equation}

Let us now revisit the scattering response of a dielectric sphere with $m=3.75$ in the guise of the squared norms of $a_1$ and $b_1$. Their corresponding line shapes in Fig.~\ref{fig:resantires}(a) show very close maxima and minima and, in particular that for $b_1$, are definitely Fano type. As can be seen, the position of fundamental antiresonances (either electric or magnetic) agrees very well with approximations with subscript $(2)$ that are marked with vertical dashed-dotted lines. Interestingly enough, it is also apparent that approximation with subscript $(0)$ is in fact much closer to the fundamental electric dipole antiresonance than to its resonant counterpart, unlike that for the magnetic dipole one (vertical solid lines). As shown in Fig.~\ref{fig:resantires}(b), there is a good agreement between $x_{res/antires(0)}^{a_1,1}$ and $x_{antires(2)}^{a_1,1}$ for all considered values of $m$. We can therefore end our discussion with the conclusion that solution of Eq.~(\ref{psi1null}), although needed in order to finally obtain an accurate approximation to $x_{res}^{a_1,1}$, does in fact provide by itself a good estimate for $x_{antires}^{a_1,1}$ rather than for the position of the fundamental electric dipole resonance.

\subsection{$\!\!\!\mathrm{\bf Approximations\ to}\ x_{res}^{a_1,j}\, \mathrm{\bf and}\ x_{res}^{b_1,j}\ \mathrm{\bf for}\ j>1, mx \gg 1$}\label{subsec:jgt1}

When considering dipole resonances with different ordinal number (that is, for $j>1$), the solutions to Eqs.~(\ref{q1null}) and (\ref{s1null}) will no longer be close to $x_{res}=1$, but may take much larger values. This implies that the condition $mx \gg 1$ can then be met for some values of $m$ that are significantly smaller than those considered in Sec.~\ref{subsec:1gg2}. In such a scenario, it seems again plausible to disregard non-sinusoidal terms in Eqs.~(\ref{psi1}) and (\ref{psi1p}) but one can not take for granted the validity of Eqs.~(\ref{q1s1bigm}). Consequently, we should keep contributions from $\chi_1\psi_1'$ and $\chi_1'\psi_1$  when defining the approximations to $q_1$ and $s_1$ for $x \gg1$:
\begin{subequations}\label{eq:q1s1xgg1}
\begin{equation}\label{q1xgg1}
q_1^{x\gg1}(x ,m)= m \cos x \cos m x +\sin x \sin mx,
\end{equation}
\begin{equation}\label{s1xgg1}
s_1^{x\gg1}(x ,m)= -m \sin x \sin m x-\cos x \cos mx.
\end{equation}
\end{subequations}
\begin{figure}
\includegraphics[width=\columnwidth]{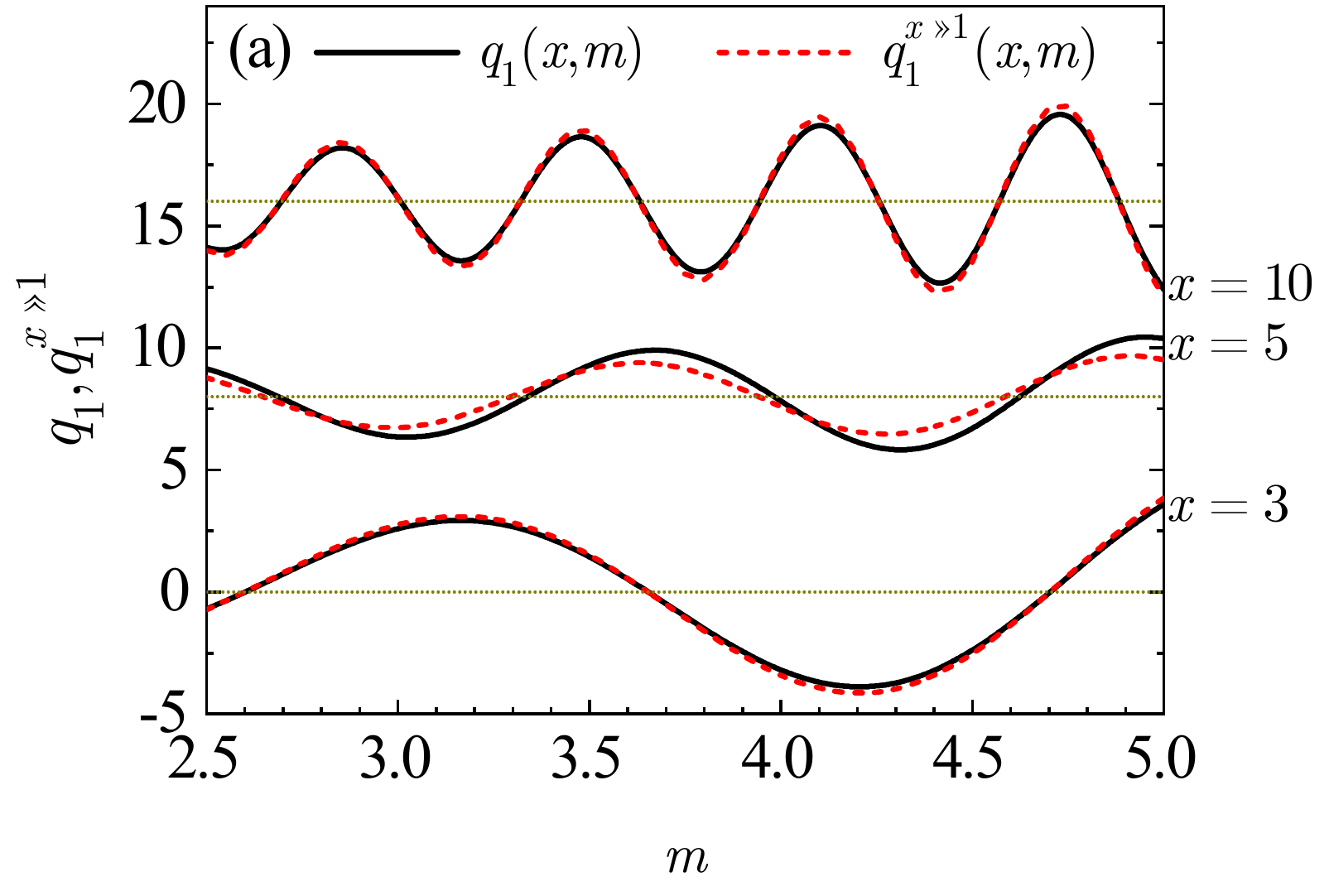}
\includegraphics[width=\columnwidth]{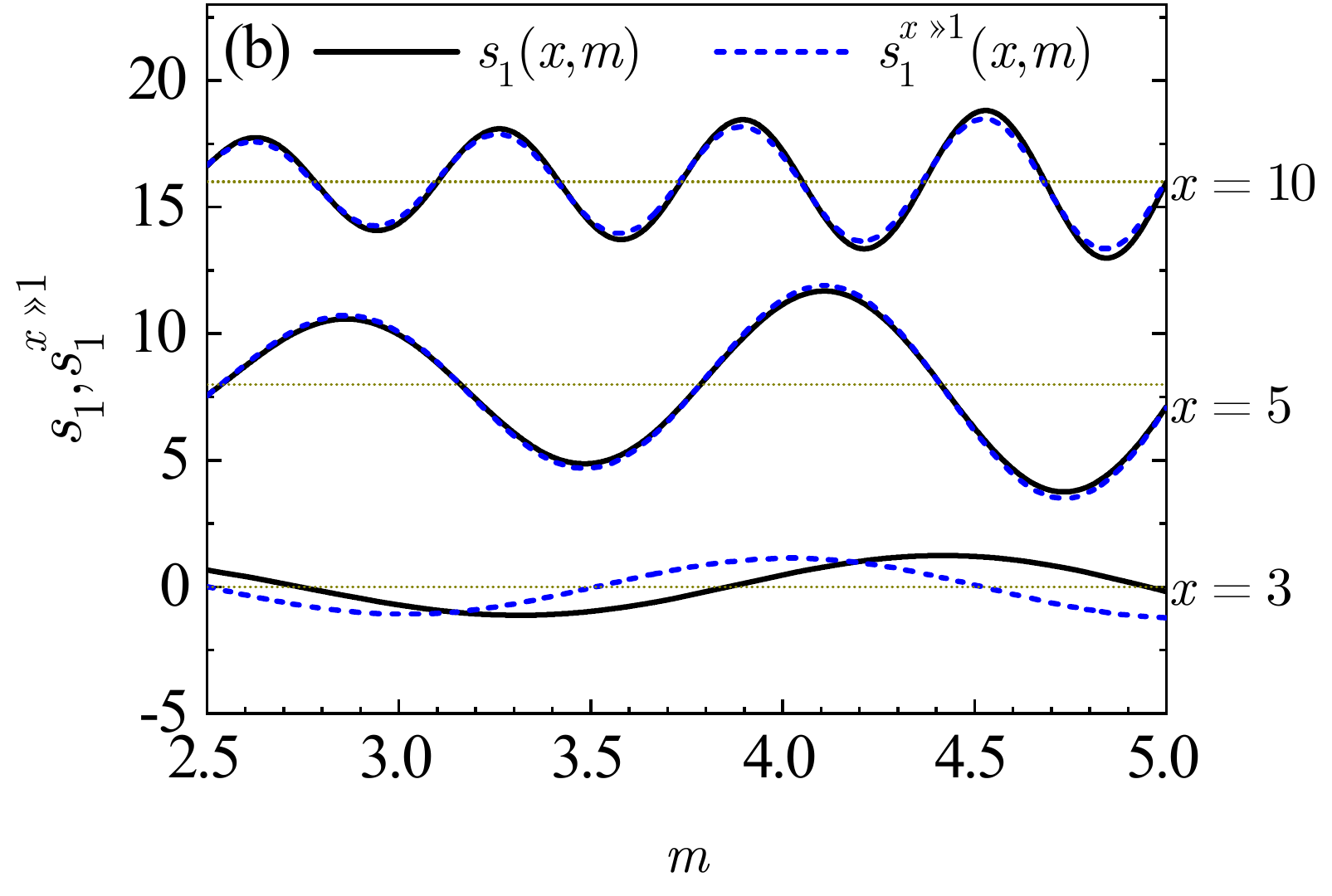}
\caption{\label{fig:q1s1vsmforx235} (a) Calculated $q_1(x,m)$ (solid) and $q_1^{x\gg1}(x,m)$ (dashed) as a function of $m$ for increasing values of size parameter. Curves corresponding to $x=3,5,10$ are artificially shifted for the sake of clarity, with their actual zero baselines marked by horizontal dotted lines. (b) Same conventions apply for $s_1(x,m)$ and $s_1^{x \gg 1}(x,m)$.}
\end{figure}

As can be seen in Fig.~\ref{fig:q1s1vsmforx235}, zeros of $q_1^{x\gg1}$  agree with those of $q_1$  for $x \gtrsim 3$ over the entire range of considered refractive index values. For the case of $s_1$ and $s_1^{x\gg1}$, such an agreement can be found for  $x \gtrsim 4$. Hence, the positive solutions to
\begin{subequations}\label{eq:q1s1xgg1null}
\begin{equation}\label{q1xgg1null}
m \cos x \cos m x +\sin x \sin mx = 0,
\end{equation}
\begin{equation}\label{s1xgg1null}
m \sin x \sin m x+\cos x \cos mx =0
\end{equation}
\end{subequations}
will provide a good approximation to successive electric and magnetic dipole resonances with $j>1$, respectively.

In order to obtain those solutions, we now return to our previous discussion of $x_{res(0)}^{a_1,j}$ and $x_{res(0)}^{b_1,j}$. For the case of electric dipole resonances, let $j>1$ and $m \gg 1$, so that the position of resonances is governed by the condition $m \cos x \cos m x =0$. Hence,
\begin{equation}\label{eq:resa1jgt1}
x_{res}^{a_1,j>1}(m)\approx x_{res^(0)}^{a_1,j>1}(m)=\frac{(j+\frac{\scriptstyle 1}{\scriptstyle 2})\pi}{m}.
\end{equation}

As $m$ goes down, $x_{res}^{a_1,j>1}$ should go up inversely, due to the continuity of size parameter and its inverse dependence on $m$. But such a continuous increase also implies that resonant $x$ should equal $(g+\tfrac{1}{2})\pi$ for some given $m$, with $g=1,2,3,\ldots$ According to Eq.~(\ref{eq:resa1jgt1}), we expect it to happen for $m=(j+\tfrac{1}{2})/(g+\tfrac{1}{2})$, but the couplet
\begin{equation}
(x,m)=\left((g+\tfrac{1}{2})\pi, \frac{(j+\tfrac{1}{2})}{(g+\tfrac{1}{2})}\right)
\end{equation}
is not a solution of Eq.~(\ref{q1xgg1null}), which is reduced to $\sin m(g+\tfrac{1}{2})\pi =0$ for $x=(g+\tfrac{1}{2})\pi$. In fact, it is  $m=(j+g)/(g+\tfrac{1}{2})$ that fulfills Eq.~(\ref{q1xgg1null}) for that particular $x$. A not-so-obvious consequence of this mathematical condition is that every time when $m$ equals $(g+j)/(g+\tfrac{1}{2})$ the resonant size parameter experiences a ``jump'' of $\tfrac{\pi}{m}$ opposite to the variation in $m$, then promoting or demoting to the adjacent zeroth-order resonance.

By simple inspection of Eq.~(\ref{q}), it is apparent that actual ``jump points'' in the size parameter of electric dipole resonances occur for those $\breve{x}_g^{a_1}$ that are solutions of $\chi_1'(\breve{x}_g^{a_1})=0$, which in fact are slightly smaller than $(g+\tfrac{1}{2})\pi$. The corresponding values of $m$ are then given by the zeros of $\psi_1'(m\breve{x}_g^{a_1})$, which can be approximated (see Appendix) by
\begin{subequations}\label{eq:mga1j}
\begin{equation}
m_g^{a_1,j}=\frac{(j+g)}{\left(g+\tfrac{1}{2}\right)}f_g^{a_1,j},
\end{equation}
with
\begin{equation}
f_g^{a_1,j}\!=\!\left[1\!+\!\frac{1}{\pi^2(g\!+\!\frac{1}{2})^2\!-\!3}\right]\!\left[1\!-\!\frac{1}{\pi^2(g\!+\!j)^2\!-\!2}\right].
\end{equation}
\end{subequations}
\begin{figure}
\includegraphics[width=\columnwidth]{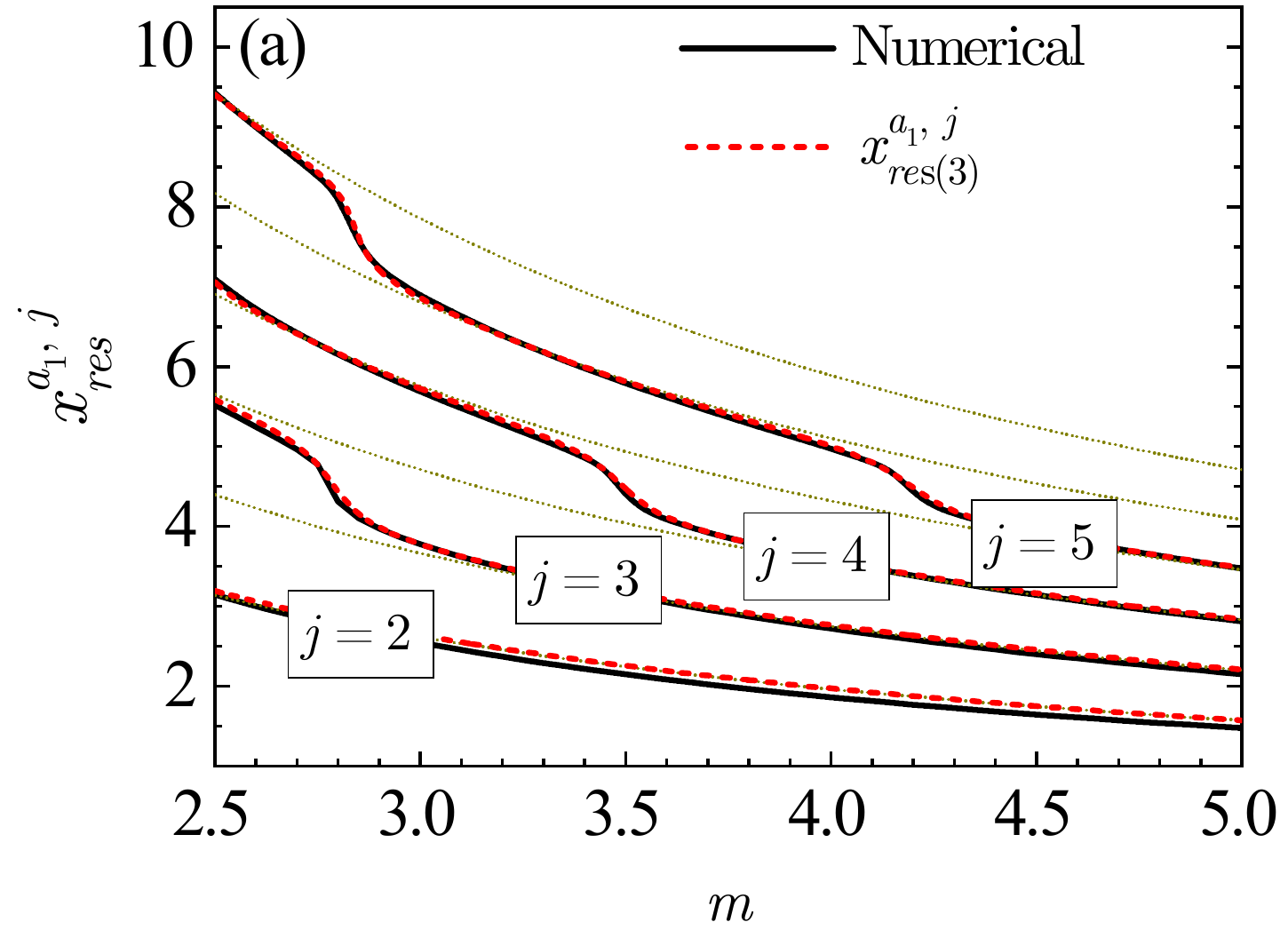}
\includegraphics[width=\columnwidth]{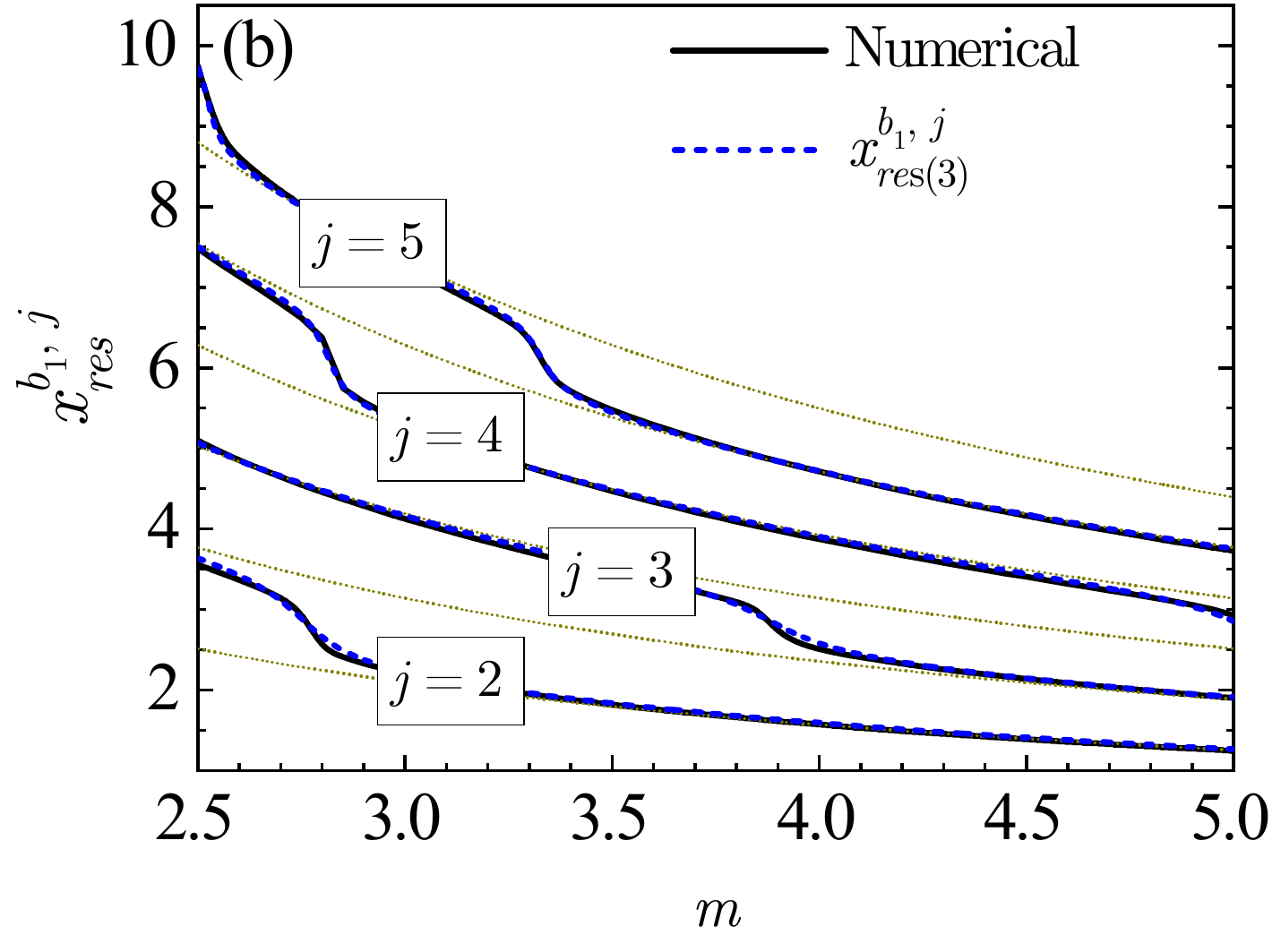}
\caption{\label{fig:xresjgt1vsm} Calculated size parameter as a function of $m$ for higher-order electric (a) and magnetic (b) dipole resonances of a non-absorbing dielectric sphere. Solid curves in (a) correspond to the numerical solutions of Eq.~(\ref{q1null}) for increasing values of $j$, whereas dashed curves show the best fits of our proposed $x_{res(3)}^{a_1,j}$ to data points (see text). Dotted curves mark the positions of sequential zeroth-order resonances. Same conventions apply for Eq.~(\ref{s1null}) and $x_{res(3)}^{b_1,j}$ in (b).}
\end{figure}
We can then expect the position of electric dipole resonances with $j=2,3, \ldots$ to be described by
\begin{equation}\label{eq:xresejgt1}
x_{res(3)}^{a_1,j}(m)\!=x_{res(0)}^{a_1,j}(m)+\frac{\pi}{m}\!\sum_{g=1}^{\infty}(1-H(m-m_g^{a_1,j})),
\end{equation}
where $H(m)$ is a smooth analytical approximation to the Heaviside step function \cite{KanwalBook}
\begin{equation}\label{eq:stepdef}
H(m)=\tfrac{1}{2}+\tfrac{1}{\pi}\arctan h m
\end{equation}
in which $h$ is left as a free parameter.

Following the same reasoning for magnetic dipole resonances (see Appendix), we obtain
\begin{equation}\label{eq:xresmjgt1}
x_{res(3)}^{b_1,j}(m)\!=x_{res(0)}^{b_1,j}(m)+\frac{\pi}{m}\!\sum_{g=1}^{\infty}(1-H(m-m_g^{b_1,j})),
\end{equation}
where values of $m$ for ``jump points'' are now given by
\begin{subequations}\label{eq:mgb1j}
\begin{equation}
m_g^{b_1,j}=\frac{\left(g+j-\frac{1}{2}\right)}{g}f_g^{b_1,j},
\end{equation}
with
\begin{equation}\label{eq:fgb1j}
\!\!f_g^{b_1,j}\!=\!\!\left[\frac{\pi^2 \!\left(g\!+\!j\!-\!\frac{1}{2}\right)^2\!\!\!-\!2}{\pi^2 \!\left(g\!+\!j\!-\!\frac{1}{2}\right)^2\!\!\!-\!1}\right]\!\!\!\left[\frac{\pi ^2 g^2\!-\!2}{2\pi ^2 g^2\!-\!3-\!\!\sqrt{\pi^4 g^4\!-\!3}}\right].\!\!\!
\end{equation}
\end{subequations}

Figure~\ref{fig:xresjgt1vsm} shows the calculated values of size parameter as a function of $m$ for successive electric and magnetic dipole resonances with $j>1$. Solid lines correspond to numerical solutions of Eqs.~(\ref{q1null}) and (\ref{s1null}), whereas dashed ones represent the best fits of expressions in Eqs.~(\ref{eq:xresejgt1}) and (\ref{eq:xresmjgt1}) to data points. Free parameter $h$ is determined for every $(j, g)$ by means of an iterative implementation of the Levenberg-Marquardt algorithm \cite{Levenberg1944, Marquardt1963}. For a given $g$, we find $h$ to be negatively proportional to $j$. On the other hand, $h$ is directly proportional to $g$ if $j$ is kept as a constant.\footnote{Obtained values for $h$ in Fig.~\ref{fig:xresjgt1vsm} can be approximated by $(2.17 g-1.11) (27.42\, -3.93 j)$ and $(2.19 g-1.49) (15.51\, -1.39 j)$ for the case of electric and magnetic dipole resonances, respectively.} As can be seen, expressions with subscript $(3)$ provide a reliable description of dipole resonances with $j>1$ for the range between $m=2.5$ and $5$, especially with respect to ``jumps'' between adjacent zeroth-order solutions. In addition, curve fitting to data points keeps the absolute percentage error below $4$ all over the considered refractive index range.\footnote{With the sole exception of $j=2$, for which the absolute percentage error reaches up to $7$ in both types of resonances.}

From Fig.~\ref{fig:xresjgt1vsm} it is also apparent that electric and magnetic dipole resonances with $j>1$ do coincide for precise values of the relative refractive index (e.g. $m\approx 2.77$ for $j=2$). However, the occurrence of ``double dipole'' resonances does not seem to cause any significant effects due to the dominance of contributions other than dipole.

\section{Dipole resonances for high- and moderate-refractive-index materials}\label{sec:dipolar}

\begin{figure}[b]
\includegraphics[width=\columnwidth]{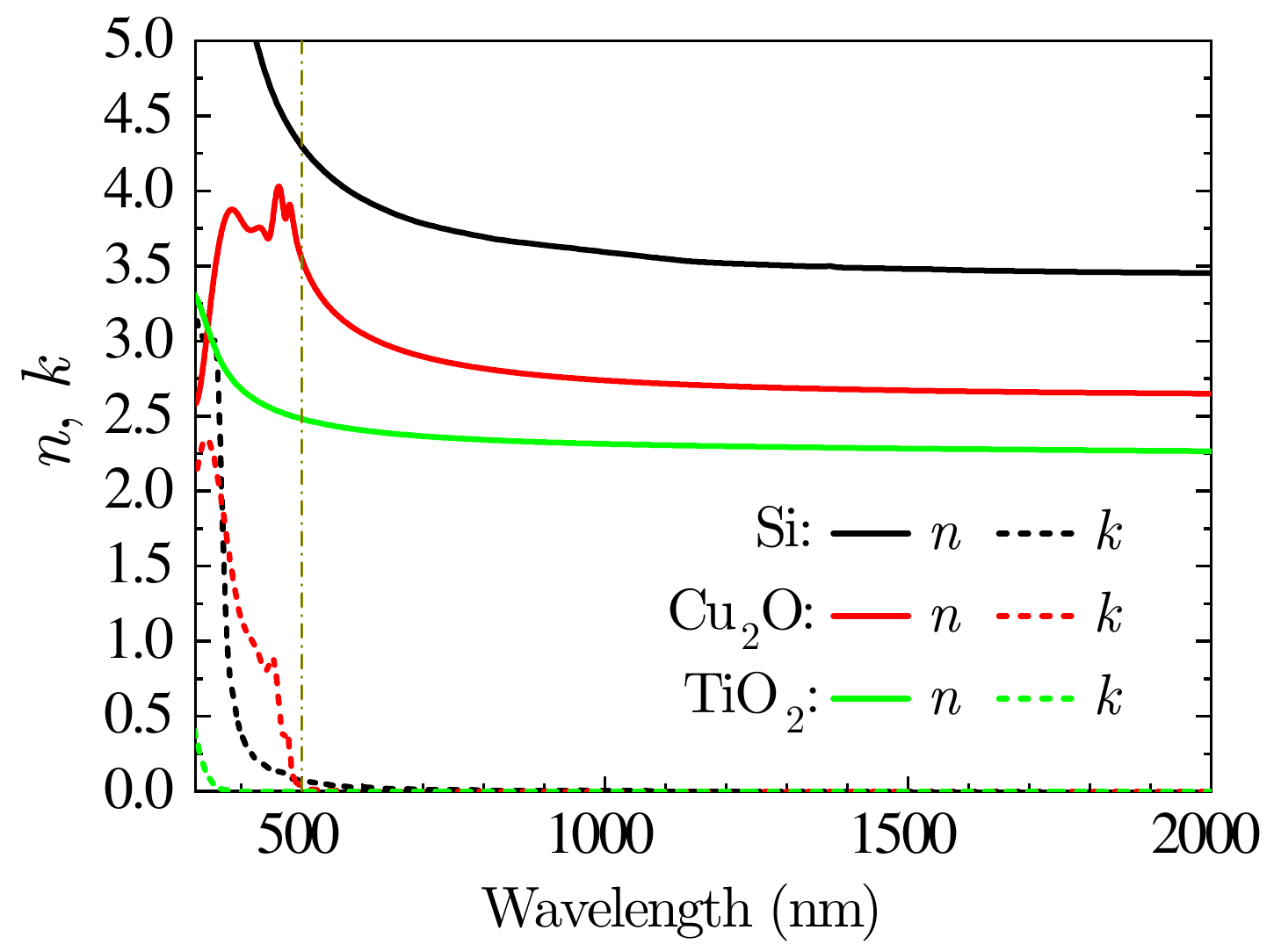}
\caption{\label{fig:rivslambda} Real (solid) and imaginary (dashed) parts of the refractive index as a function of wavelength for Si, $\mathrm{Cu_2O}$ and $\mathrm{TiO_2}$, obtained from Refs.~\onlinecite{Green2008, Haidu2011, Siefke2016}, respectively. Vertical dashed-dotted line marks the wavelength from which absorption can be neglected.}
\end{figure}

\begin{figure*}
\includegraphics[width=\columnwidth]{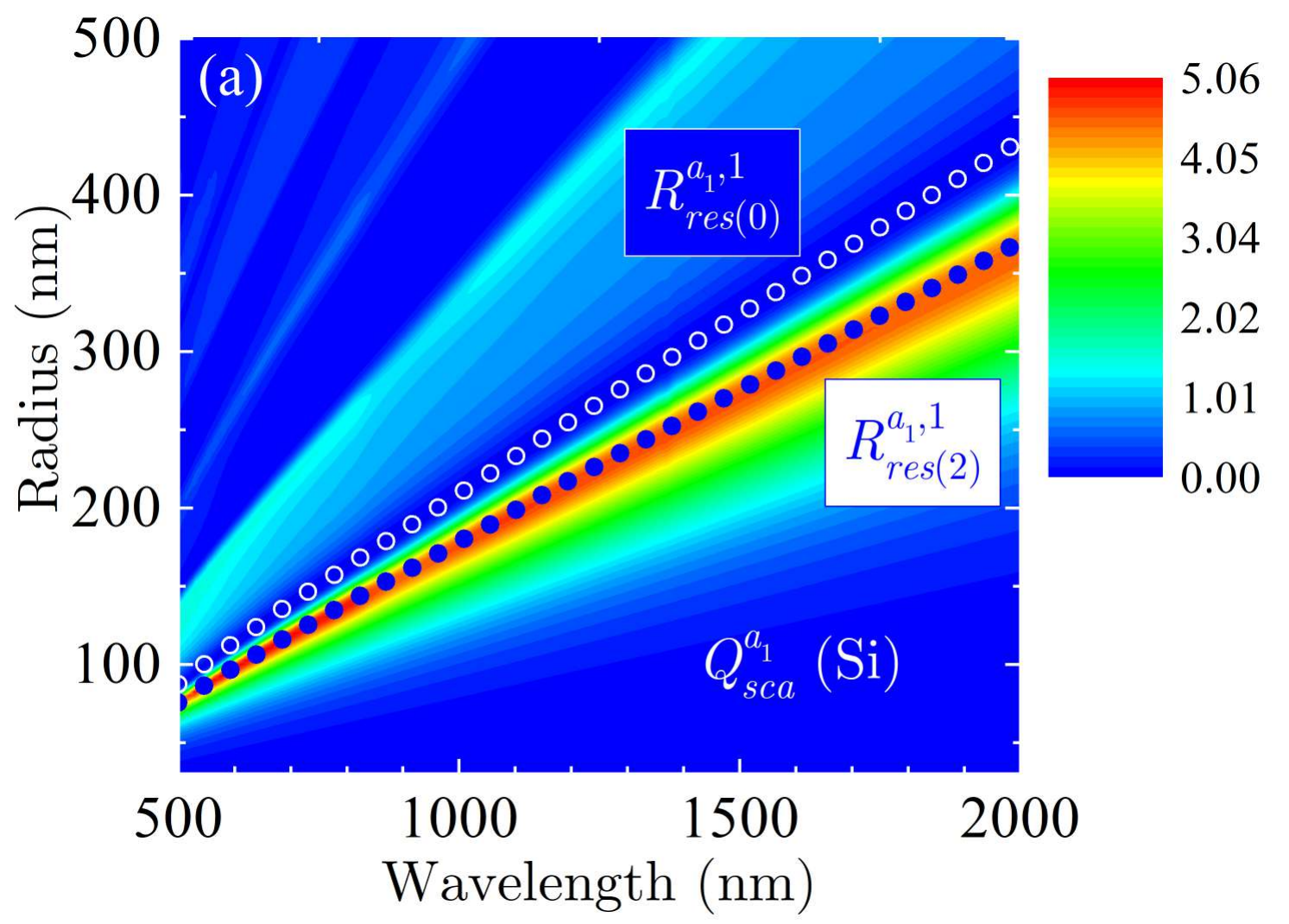}
\includegraphics[width=\columnwidth]{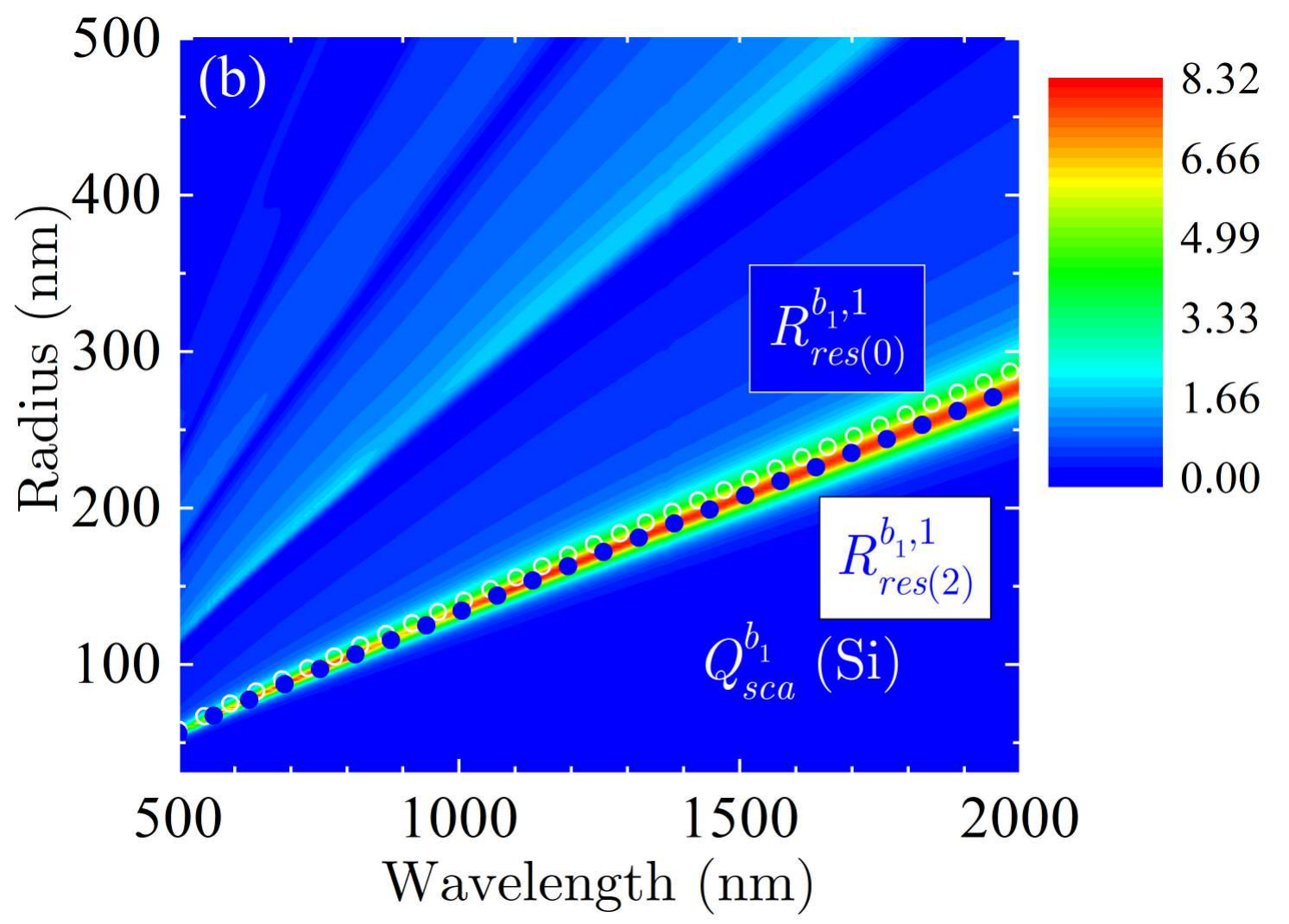}
\includegraphics[width=\columnwidth]{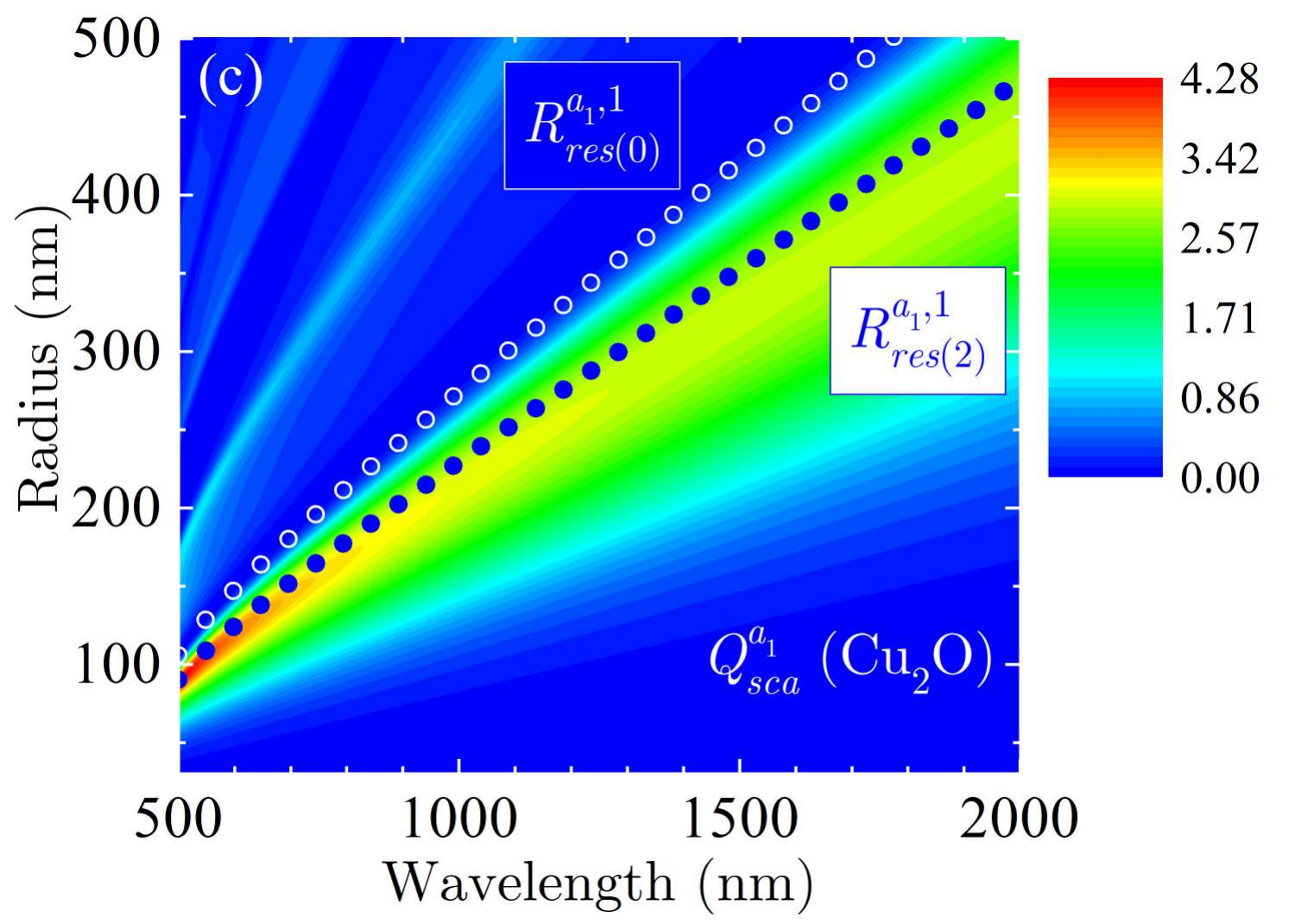}
\includegraphics[width=\columnwidth]{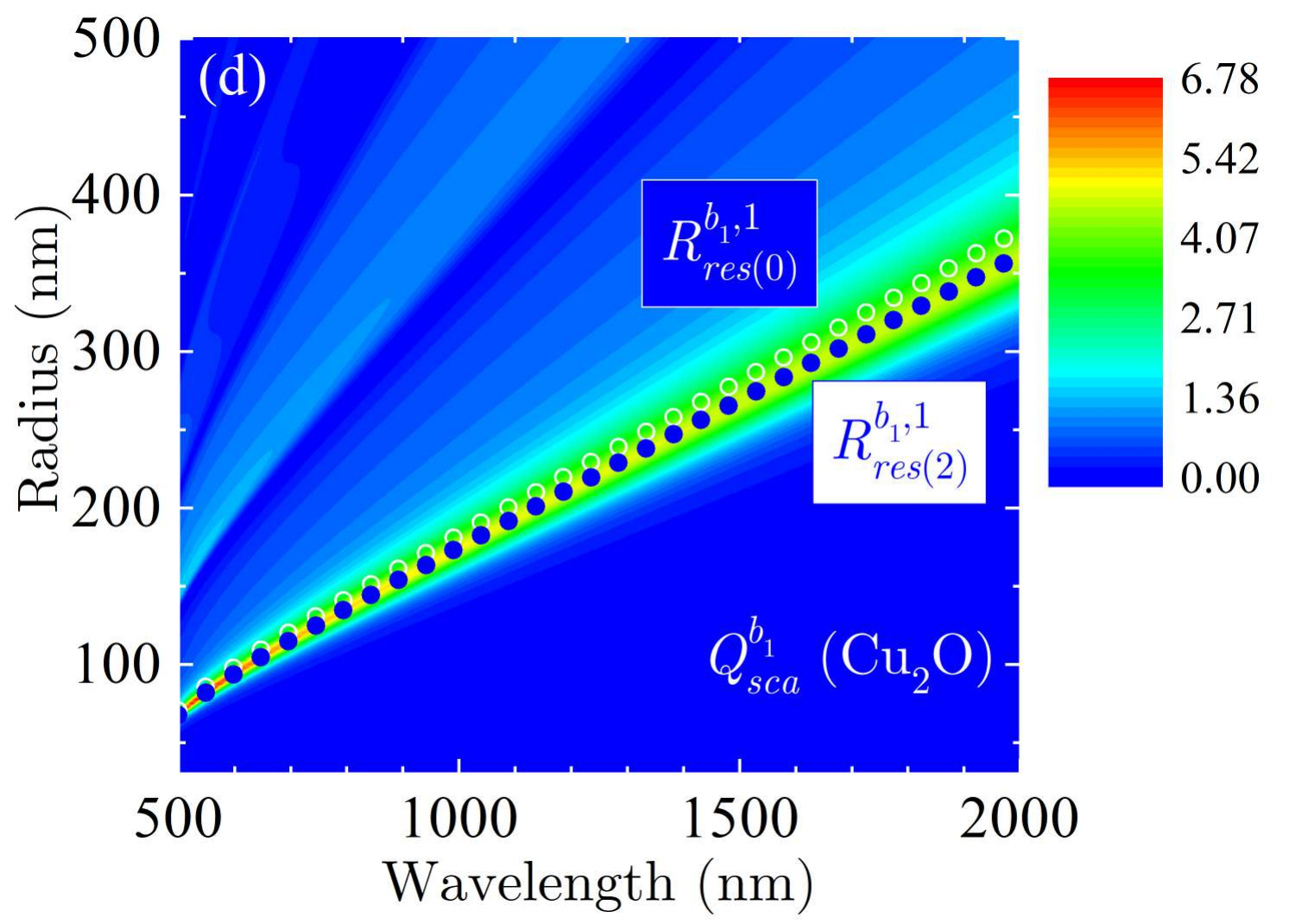}
\includegraphics[width=\columnwidth]{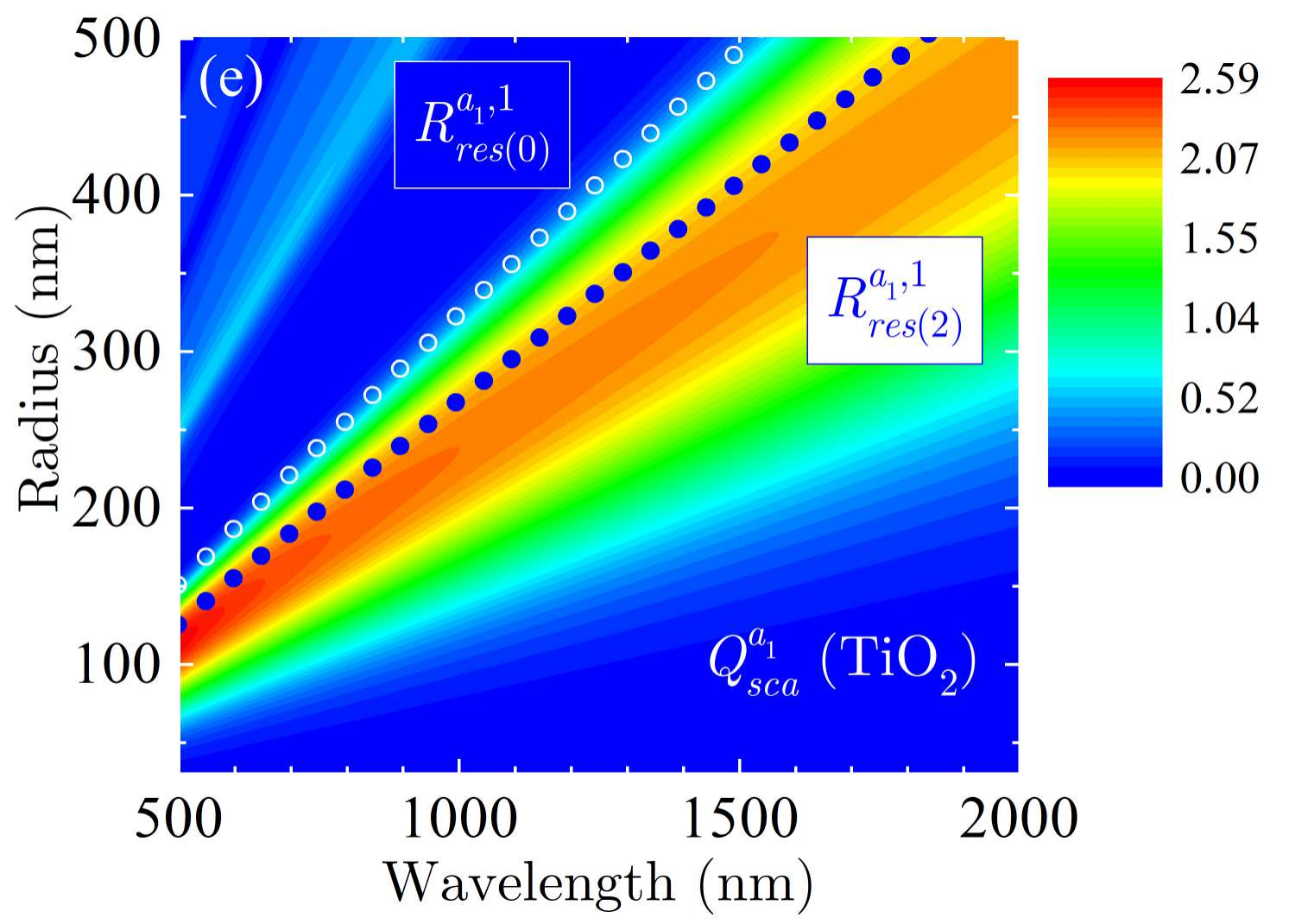}
\includegraphics[width=\columnwidth]{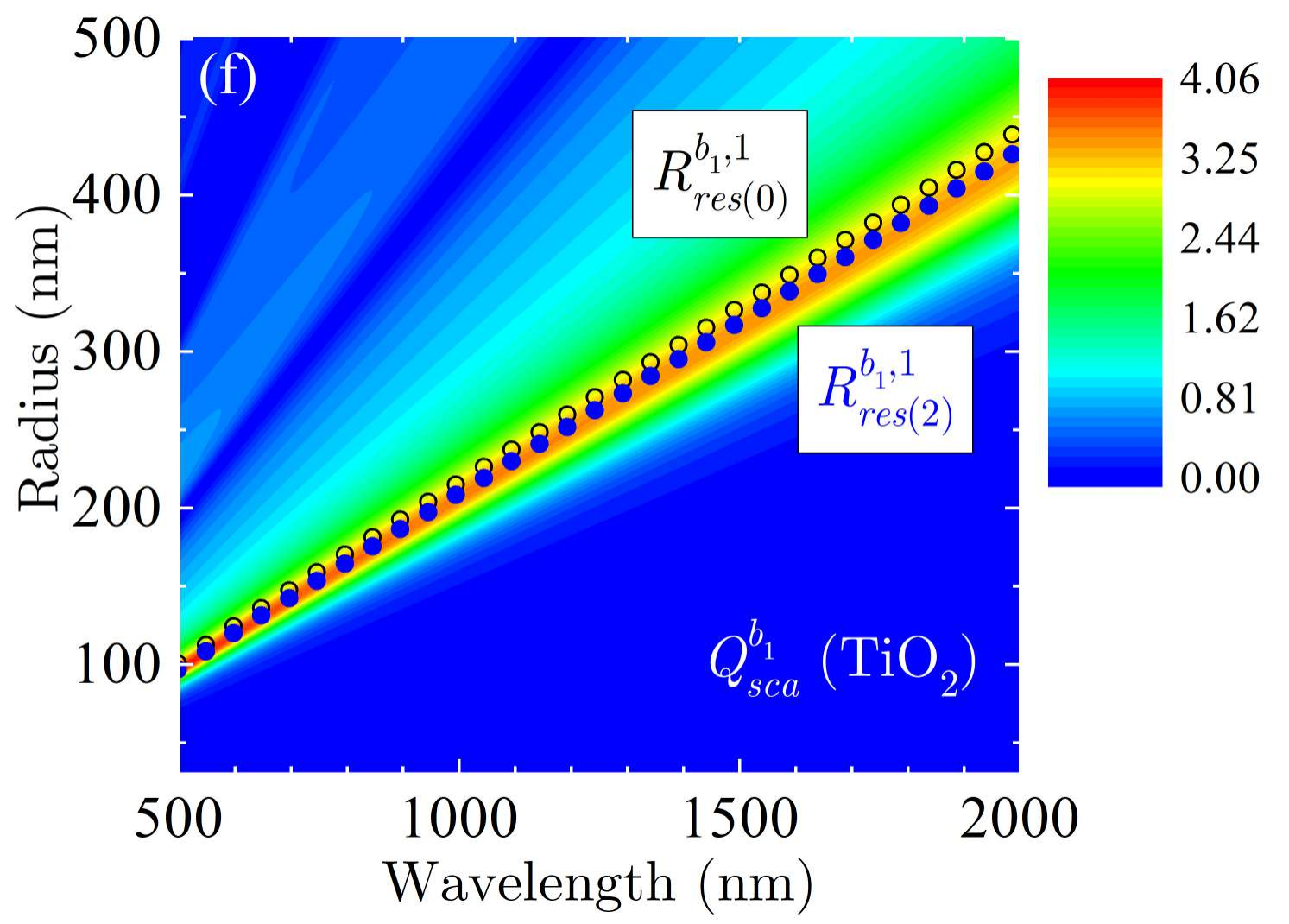}
\caption{\label{fig:cps} Electric and magnetic dipole contributions to the scattering efficiency of a dielectric sphere as a function of the incident wavelength and the sphere's radius for three different materials: Si (a,b), $\mathrm{Cu_2O}$ (c,d) and $\mathrm{TiO_2}$ (e,f). Open ($\circ$) and solid ($\bullet$) symbols correspond to estimates for resonant radii with subscripts $(0)$ and $(2)$ respectively.}
\end{figure*}

Up to this point, we have been focused on the solution of equations. We now return to the scattering properties of actual dielectric nanospheres in the optical range. For the sake of simplicity, let us assume that our sphere is surrounded by air, so that we can replace $m$ by the sphere's complex refractive index $n+\it{i}k$. Given that all our findings have been obtained for non-absorbing materials, we require $k \approx 0$. In Fig.~\ref{fig:rivslambda} we present the real and imaginary parts of the refractive index as a function of wavelength for Si, $\mathrm{Cu_2O}$, and $\mathrm{TiO_2}$, according to Refs.~\onlinecite{Green2008, Haidu2011, Siefke2016}, respectively. As can be seen, these three materials fulfill the requirement of not absorbing light within the interval between 500 and 2000~nm and have therefore been the subject of recent experimental research on dielectric nanoresonators \cite{Evlyukhin2012, Kuznetsov2012, Zhang2015, Barreda2017, Susman2017, Ullah2018, Yavas2019, Zhuo2019}. In addition, the range of values of $n$ within such interval for Si, $\mathrm{Cu_2O}$ and $\mathrm{TiO_2}$ cover most of that of $m$ that was considered in the previous section. These materials seem, then, to be convenient to test our improved approximate conditions by comparing their predicted resonances with those obtained from the full calculation of $Q_{sca}^{a_1}$ and
$Q_{sca}^{b_1}$.

\begin{figure*}
\includegraphics[width=2\columnwidth]{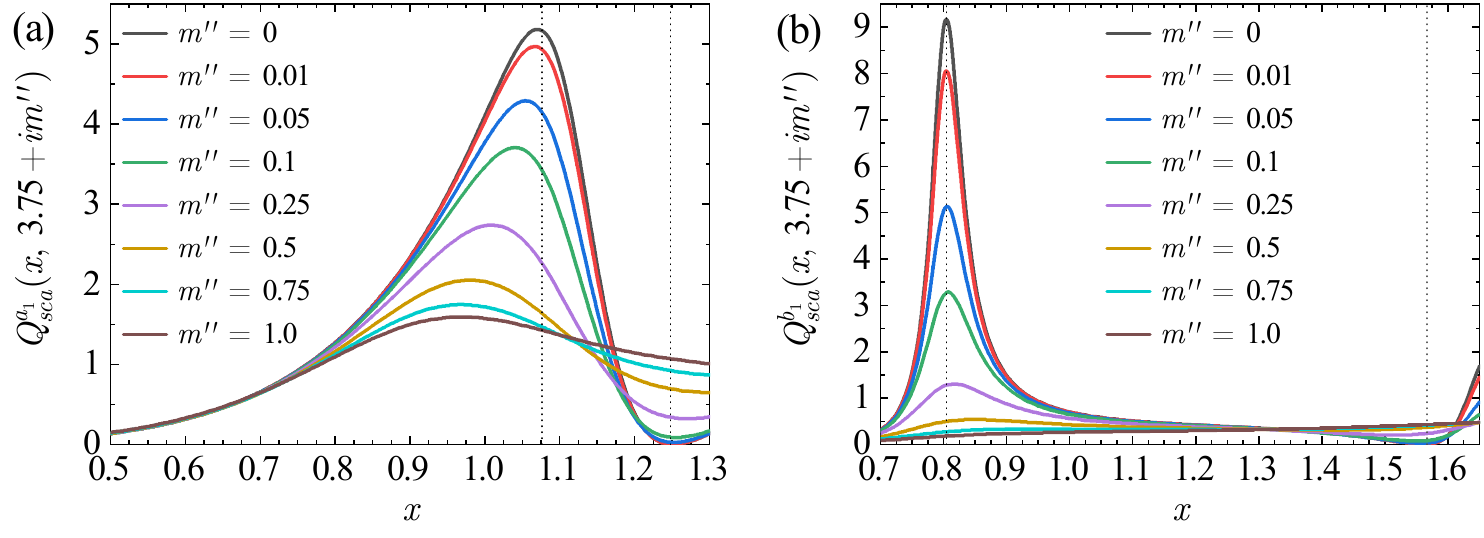}
\caption{\label{fig:abs} Electric (a) and magnetic (b) dipole contributions to the scattering efficiency as a function of size parameter $x$ for a dielectric sphere with $m=3.75 +im''$. Solid curves correspond to increasing values of $m''$. Vertical dashed lines mark the approximated positions of fundamental dipole resonances and antiresonances for $m=3.75$.}
\end{figure*}

It follows from the very definition of size parameter that
\begin{equation}\label{eq:R_res}
R_{res}^{\square,1}=\frac{\displaystyle \lambda}{\displaystyle 2 \pi}  x_{res}^{\square,1}(m (\lambda)),
\end{equation}
where $\square$ stands for either $a_1$ or $b_1$. We then substitute for $x_{res}^{\square,1}$ from Eqs.~(\ref{eq:xa112}) and (\ref{eq:xb112}) into Eq.~(\ref{eq:R_res}) in order to obtain the best estimates for $(m, \lambda, R)$ triplets exhibiting fundamental dipole resonances. For the sake of comparison, we also calculate the resonant radii corresponding to approximations with subscript $(0)$.

Panels in Fig.~\ref{fig:cps} show the calculated electric and magnetic dipole contributions to the scattering efficiency of a dielectric sphere as a function of the incident wavelength and the sphere's radius for Si, $\mathrm{Cu_2O}$ and $\mathrm{TiO_2}$. All calculated values result from straightforward evaluation of $Q_{sca}^{a_1}$ and $Q_{sca}^{b_1}$ by means of a homemade \emph{Mathematica} script.\footnote{Wolfram Research, Inc., Mathematica, Version 9.0, Champaign, IL (2012)} Open ($\circ$) and solid ($\bullet$) symbols correspond to the above-mentioned $R_{res(0)}^{\square,1}$ and $R_{res(2)}^{\square,1}$ respectively.

Prior to discussing results in Fig.~\ref{fig:cps}, we have to keep in mind that the positions of resonances for $|\square|^2$ do not exactly coincide with those for $Q_{sca}^{\square}$ because of the extra $x^{-2}$ factor that appears in the definition of dipole scattering efficiencies (see Sec.~\ref{sec:lightscs}). As a consequence of the different symmetry of their Fano-like line profiles (see for example, those in Fig.~\ref{fig:resantires}(a)), such discrepancy becomes more apparent for the fundamental electric dipole resonance than for the fundamental magnetic one. This means that, when plotted against $\lambda$, calculated $Q_{sca}^{a_1}$ for a given sphere radius reaches its maximum at a wavelength that is always red-shifted with respect to $2 \pi R/x_{res}^{a_1,1}$. The wavelength shift is inversely proportional to the resonant value for $m$. Aside from these subtleties, which are particularly visible for the case of titanium dioxide in Fig.~\ref{fig:cps}(e), there is an excellent agreement between estimates to resonant radii and calculated dipole efficiencies for all three materials across the considered wavelength range. In addition, it is apparent that, by increasing the wavelength, estimates with subscript $(0)$ depart from the actual resonances exactly in the same fashion as they do $x_{res(0)}^{a_1,1}$ and $x_{res(0)}^{b_1,1}$ when $m$ decreases (see Fig.~\ref{fig:xresvsm}).

In fact, Fig.~\ref{fig:cps} brings us up against some physical interpretation of resonant $(m, \lambda, R)$ triplets. From considerations based on geometrical optics (see, for example, Ref.~\onlinecite{Roll2000}), it can be figured out that electric dipole resonances occur when $2R$ becomes approximately equal to an odd multiple of the half-wavelength inside the sphere, which is precisely the prediction of Eq.~(\ref{eq:xa1_0}). However, Figs.~\ref{fig:cps}(a), \ref{fig:cps}(c) and \ref{fig:cps}(e) show that, as $m$ decreases, electric dipole resonances take place for radii that are smaller than $R_{res(0)}^{a_1,1}$, thus pointing out some sort of effective increasing of the sphere's size for moderate values of $m$. In contrast, there is no such re-sizing for magnetic dipole resonances, which appear for diameters that are equal to an integer multiple of $\lambda/m$, aside from the correction prescribed by Eq.~({\ref{deltaxb10}). Finally, signatures of resonances with $j=2$ are clearly apparent in the upper left quadrant of every panel in Fig.~\ref{fig:cps}. Nevertheless, their corresponding scattering efficiencies are about one fifth of those of the fundamental resonances and they do not seem to be especially relevant for any of these materials.

Given that the zero-absorption threshold defined in Fig.~\ref{fig:rivslambda} is somewhat arbitrary, we cannot close this section without discussing the robustness of our obtained approximations when dealing with some degree of dissipation. For a complex relative refractive index $m=m'+im''$, there is no unequivocal correspondence between resonances and antiresonances in $Q_{sca}^{a_1}$ and $Q_{sca}^{b_1}$ and zeros and poles in $a_1$ and $b_1$, so that explicit expressions for resonant or antiresonant values of size parameter become practically unattainable. However, one could expect approximations based on the real part of $m$ to still hold for a weakly absorbing medium.

As a test for this hypothesis, we present in Fig.~\ref{fig:abs} the calculated values of electric and magnetic dipole contributions to the scattering efficiency as a function of size parameter $x$ for a dielectric sphere with $m=3.75 +im''$. Such a fixed value for $m'$ corresponds to the midpoint of the interval between $2.5$ and $5$ that has been considered all along this work. It is also in the range between those of $n$ for Si and $\mathrm{Cu_2O}$ at the wavelength from which absorption seems negligible in Fig.~\ref{fig:rivslambda}. With respect to $m''$, it is gradually increased from $0$ to $1$, which is the maximum value of $\kappa$ for silicon and cuprous oxide above $400$~nm. Please keep in mind that we have chosen this setting for testing purposes only, as far as $m''$ is connected with $m'$ by causality and cannot therefore take arbitrary values.

As can be seen, all spectral features are significantly damped and also slightly shifted as dissipation increases. Direction of the spectral shift with respect to approximations with subscript $(2)$ for $m=3.75$ (vertical dashed lines) seems to be opposite for fundamental electric and magnetic resonances and antiresonances. Thus, the position of fundamental electric dipole resonance shifts from $x=1.07$ for $m''=0$ to $x=0.97$ ($-9\%$) for $m''=1.0$. In contrast, $x_{antires}^{a_1,1}$ shifts oppositely ($+5\%$) for $m''=0.5$, which is the highest of the considered values that permits the resolution of the dip. On the other hand, the size parameter of fundamental magnetic dipole resonance evolves from $x=0.805$ for $m''=0$ to $x=0.85$ ($+6\%$) for $m''=0.5$, whereas that for the antiresonance does from $x=1.56$ to $x=1.425$ ($-9\%$). We can then conclude that Eqs.~(\ref{eq:xa112}), (\ref{eq:xb112}), (\ref{eq:xantia12}), and (\ref{eq:xantib112}) can be reasonably extended to the entire visible range for Si, $\mathrm{Cu_2O}$ and $\mathrm{TiO_2}$, which seems convenient for designing purposes.

\section{Conclusions}\label{sec:conclu}

We have obtained explicit expressions that provide accurate approximations to dipole resonances and antiresonances in the scattering spectrum of non-absorbing dielectric nanospheres with high- and moderate-refractive-index values in the optical range. These expressions enable us to predict the occurrence of a dipole resonance with any ordinal number for a triplet of sphere's radius $R$, incident wavelength $\lambda$ and relative refractive index value $m$ without the actual evaluation of Mie scattering coefficients. Our predictions retrieve previous results for $m \gg 1$ and extend them to a wider range. We have confirmed their validity for specific dielectric materials that are widely used in photonic devices. Therefore, we expect our results to be useful for the designing of dielectric nanoresonators, particularly for issues such as biosensing \cite{Yavas2019}, nanoscopy \cite{Lukyanchuk2017} or photonic nanojet lithography \cite{DeepakKallepalli2013}.

\begin{acknowledgments}
I would like to thank A.~I. Fern\'{a}ndez-Dom\'{\i}nguez, C. Gonz\'{a}lez, M. Rey, D. Vald\'{e}s, and H. V\'{a}zquez for valuable inputs and discussions. This research work is co-funded by Gobierno de Arag\'{o}n (Grant No. $E10\_17D$) and Programa Operativo FEDER Arag\'{o}n 2014-2020 ``Construyendo Europa desde Arag\'{o}n''.
\end{acknowledgments}

\appendix*
\section{DETERMINATION OF $m_g^{a_1,j}, m_g^{b_1,j}$}\label{Appg1}

As stated in Sec.~\ref{subsec:jgt1}, ``jump points'' in the size parameter of electric dipole resonances occur for those $\breve{x}_g^{a_1}$ that are solutions of $\chi_1'(\breve{x}_g^{a_1})=0$ with $g=1,2,3,\ldots$. For all practical purposes, we can define $\breve{x}_g^{a_1}$ as the zeros of the linear Taylor expansion of $\chi_1'(x)$ about $x=(g+\tfrac{1}{2})\pi$:
\begin{equation}\label{eq:xjumpa1g}
\breve{x}_g^{a_1}\!=\!\left(g+\tfrac{1}{2}\right)\!\pi\!\!\left[\frac{\pi^2(1+2g)^2-12}{\pi^2(1+2g)^2-8}\right]\! \lesssim \left(g+\tfrac{1}{2}\right)\!\pi.
\end{equation}
The corresponding values of $m_g^{a_1,j}$ are the solutions of $q_1(\breve{x}_g^{a_1},m)=0$, which do coincide with those of $\psi_1'(m\breve{x}_g^{a_1})=0$ (see Fig.~\ref{fig:q1s1vsmforPi}(a)). Such solutions can be very well approximated by the zeros of the linear Taylor expansion of $\psi_1'(m\breve{x}_g^{a_1})$ about $m=(j+g)\pi/\breve{x}_g^{a_1}$ that are presented in Eqs.~(\ref{eq:mga1j}). For $g=1$, $\breve{x}_1^{a_1}\approx 0.95\left(\tfrac{3}{2}\pi\right)$ and the values of $m_1^{a_1,j}$ are slightly greater than $\tfrac{2}{3}(j+1)$, as shown by vertical dotted lines in Fig.~\ref{fig:q1s1vsmforPi}(a):
\begin{equation}\label{eq:num1a1j}
m_1^{a_1,\{3,4,5,6\ldots\}}=\{2.787, 3.493, 4.196, 4.899\ldots\}.
\end{equation}

\begin{figure*}[t]
\includegraphics[width=\columnwidth]{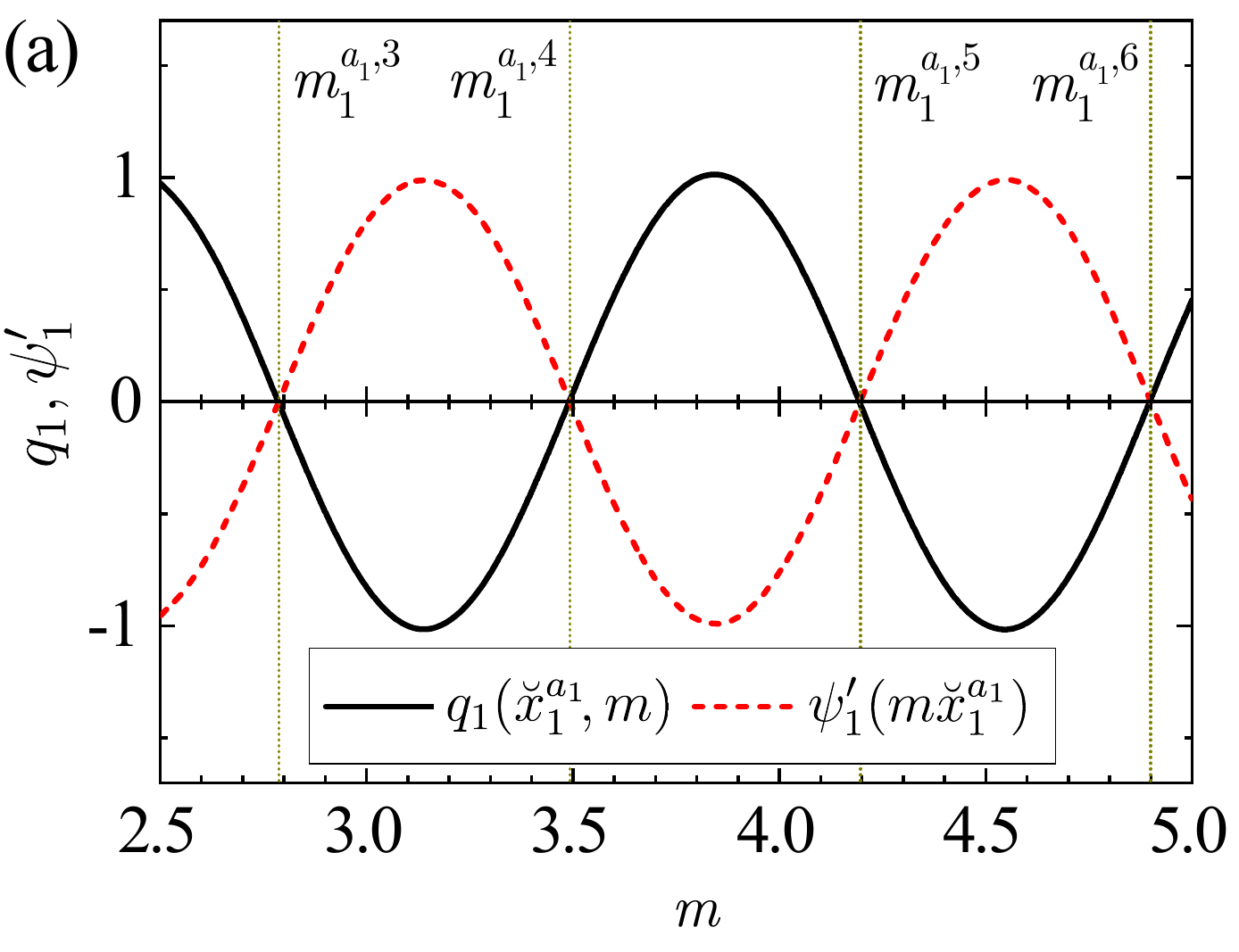}
\includegraphics[width=\columnwidth]{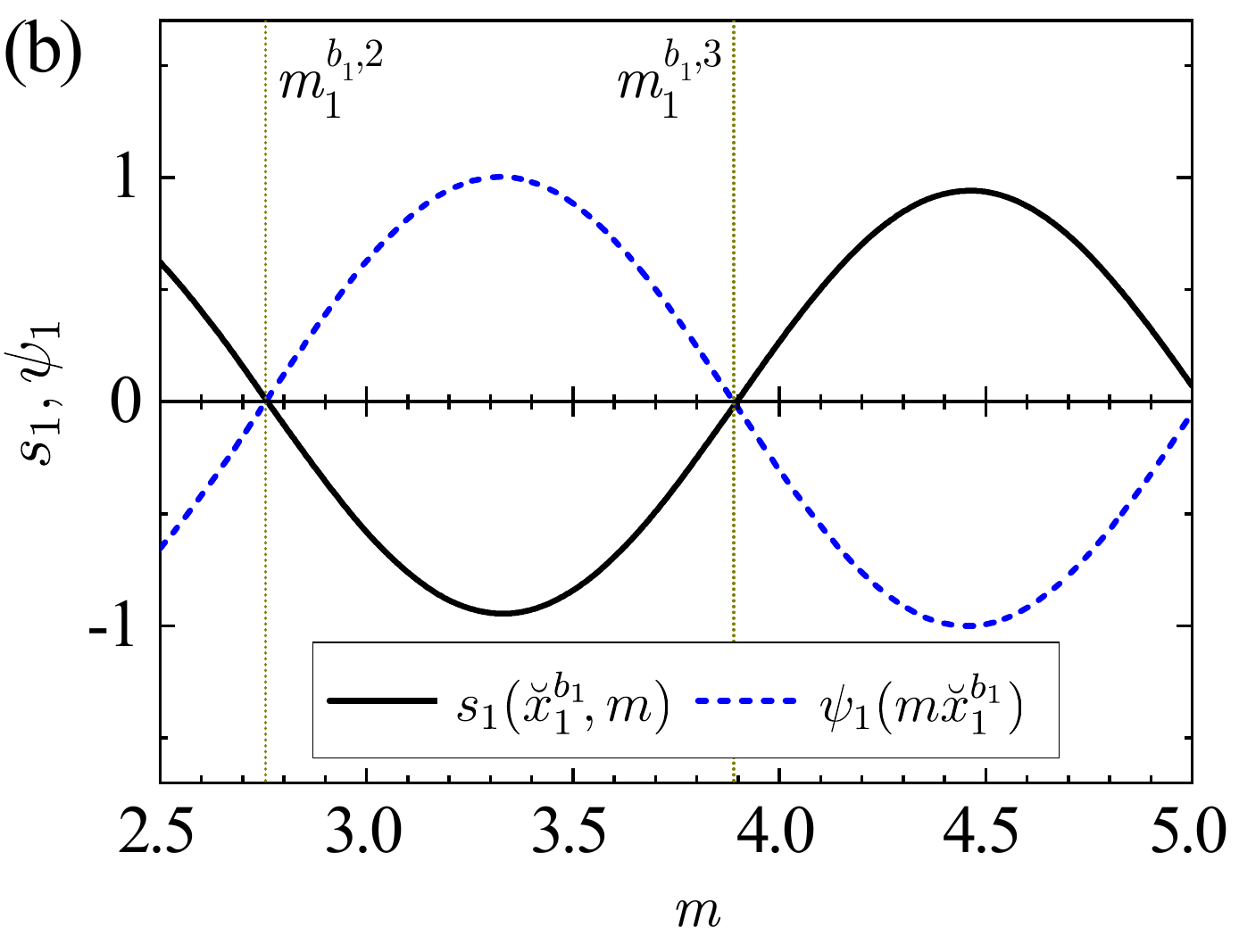}
\caption{\label{fig:q1s1vsmforPi} (a) Calculated $q_1(\breve{x}_1^{a_1},m)$ (solid) and $\psi_1'(m\breve{x}_1^{a_1}))$ (dashed) as a function of $m$. Vertical dotted lines mark the values of $m_1^{a_1,j}$ from \ref{eq:num1a1j}. (b) Calculated $s_1(\breve{x}_1^{b_1},m)$ (solid), $\psi_1(m\breve{x}_1^{b_1})$ (dashed) as a function of $m$. Vertical dotted lines mark the values of $m_1^{b_1,j}$ from \ref{eq:num1b1j}.}
\end{figure*}
For magnetic dipole resonances, ``jump points'' appear for those $\breve{x}_g^{b_1}$ that are solutions of $\chi_1(\breve{x}_g^{b_1})=0$. In contrast to dielectric ones, we have defined $\breve{x}_g^{b_1}$ as the zeros of the quadratic Taylor expansion of $\chi_1(x)$ about $x=g\pi$ in order to improve their numerical accuracy:
\begin{equation}\label{eq:xjumpb1g}
\breve{x}_g^{b_1}=g\pi\left[2-\frac{\sqrt{g^4\pi^4-3}-1}{g^2\pi^2-2}\right] \lesssim g\pi.
\end{equation}
The corresponding values of $m_g^{b_1,j}$ are the solutions of $s_1(\breve{x}_g^{b_1},m)=0$, which do coincide with those of $\psi_1(m\breve{x}_g^{b_1})=0$ (see Fig.~\ref{fig:q1s1vsmforPi}(b)). Fortunately, those solutions can be very well approximated by the zeros of the linear Taylor expansion of $\psi_1(m\breve{x}_g^{b_1})$ about $m=(j+g-\tfrac{1}{2})\pi/\breve{x}_g^{b_1}$ that are presented in Eqs.~(\ref{eq:mgb1j}). For $g=1$, $\breve{x}_1^{b_1}\approx 0.89\pi$ and the values of $m_1^{b_1,j}$ are significantly greater than $(j+\tfrac{1}{2})$, as shown by vertical dotted lines in Fig.~\ref{fig:q1s1vsmforPi}(b):
\begin{equation}\label{eq:num1b1j}
m_1^{b_1,\{2,3,4,\ldots\}}=\{2.755,3.889,5.017,\ldots\}.
\end{equation}

\bibliography{paper_a1b1_resub3}

\begin{thebibliography}{50}%
\makeatletter
\providecommand \@ifxundefined [1]{%
 \@ifx{#1\undefined}
}%
\providecommand \@ifnum [1]{%
 \ifnum #1\expandafter \@firstoftwo
 \else \expandafter \@secondoftwo
 \fi
}%
\providecommand \@ifx [1]{%
 \ifx #1\expandafter \@firstoftwo
 \else \expandafter \@secondoftwo
 \fi
}%
\providecommand \natexlab [1]{#1}%
\providecommand \enquote  [1]{``#1''}%
\providecommand \bibnamefont  [1]{#1}%
\providecommand \bibfnamefont [1]{#1}%
\providecommand \citenamefont [1]{#1}%
\providecommand \href@noop [0]{\@secondoftwo}%
\providecommand \href [0]{\begingroup \@sanitize@url \@href}%
\providecommand \@href[1]{\@@startlink{#1}\@@href}%
\providecommand \@@href[1]{\endgroup#1\@@endlink}%
\providecommand \@sanitize@url [0]{\catcode `\\12\catcode `\$12\catcode
  `\&12\catcode `\#12\catcode `\^12\catcode `\_12\catcode `\%12\relax}%
\providecommand \@@startlink[1]{}%
\providecommand \@@endlink[0]{}%
\providecommand \url  [0]{\begingroup\@sanitize@url \@url }%
\providecommand \@url [1]{\endgroup\@href {#1}{\urlprefix }}%
\providecommand \urlprefix  [0]{URL }%
\providecommand \Eprint [0]{\href }%
\providecommand \doibase [0]{http://dx.doi.org/}%
\providecommand \selectlanguage [0]{\@gobble}%
\providecommand \bibinfo  [0]{\@secondoftwo}%
\providecommand \bibfield  [0]{\@secondoftwo}%
\providecommand \translation [1]{[#1]}%
\providecommand \BibitemOpen [0]{}%
\providecommand \bibitemStop [0]{}%
\providecommand \bibitemNoStop [0]{.\EOS\space}%
\providecommand \EOS [0]{\spacefactor3000\relax}%
\providecommand \BibitemShut  [1]{\csname bibitem#1\endcsname}%
\let\auto@bib@innerbib\@empty
\bibitem [{\citenamefont {Bharadwaj}\ \emph {et~al.}(2009)\citenamefont
  {Bharadwaj}, \citenamefont {Deutsch},\ and\ \citenamefont
  {Novotny}}]{Bharadwaj2009}%
  \BibitemOpen
  \bibfield  {author} {\bibinfo {author} {\bibfnamefont {P.}~\bibnamefont
  {Bharadwaj}}, \bibinfo {author} {\bibfnamefont {B.}~\bibnamefont {Deutsch}},
  \ and\ \bibinfo {author} {\bibfnamefont {L.}~\bibnamefont {Novotny}},\
  }\bibfield  {title} {\enquote {\bibinfo {title} {{Optical Antennas}},}\
  }\href@noop {} {\bibfield  {journal} {\bibinfo  {journal} {Adv. Opt.
  Photonics}\ }\textbf {\bibinfo {volume} {1}},\ \bibinfo {pages} {438}
  (\bibinfo {year} {2009})}\BibitemShut {NoStop}%
\bibitem [{\citenamefont {Soukoulis}\ \emph {et~al.}(2007)\citenamefont
  {Soukoulis}, \citenamefont {Linden},\ and\ \citenamefont
  {Wegener}}]{Soukoulis2007}%
  \BibitemOpen
  \bibfield  {author} {\bibinfo {author} {\bibfnamefont {C.~M.}\ \bibnamefont
  {Soukoulis}}, \bibinfo {author} {\bibfnamefont {S.}~\bibnamefont {Linden}}, \
  and\ \bibinfo {author} {\bibfnamefont {M.}~\bibnamefont {Wegener}},\
  }\bibfield  {title} {\enquote {\bibinfo {title} {{Negative refractive index
  at optical wavelengths}},}\ }\href@noop {} {\bibfield  {journal} {\bibinfo
  {journal} {Science}\ }\textbf {\bibinfo {volume} {315}},\ \bibinfo {pages}
  {47--49} (\bibinfo {year} {2007})}\BibitemShut {NoStop}%
\bibitem [{\citenamefont {Dintinger}\ \emph {et~al.}(2012)\citenamefont
  {Dintinger}, \citenamefont {M{\"{u}}hlig}, \citenamefont {Rockstuhl},\ and\
  \citenamefont {Scharf}}]{Dintinger2012}%
  \BibitemOpen
  \bibfield  {author} {\bibinfo {author} {\bibfnamefont {J.}~\bibnamefont
  {Dintinger}}, \bibinfo {author} {\bibfnamefont {S.}~\bibnamefont
  {M{\"{u}}hlig}}, \bibinfo {author} {\bibfnamefont {C.}~\bibnamefont
  {Rockstuhl}}, \ and\ \bibinfo {author} {\bibfnamefont {T.}~\bibnamefont
  {Scharf}},\ }\bibfield  {title} {\enquote {\bibinfo {title} {{A bottom-up
  approach to fabricate optical metamaterials by self-assembled metallic
  nanoparticles}},}\ }\href@noop {} {\bibfield  {journal} {\bibinfo  {journal}
  {Opt. Mater. Express}\ }\textbf {\bibinfo {volume} {2}},\ \bibinfo {pages}
  {269--278} (\bibinfo {year} {2012})}\BibitemShut {NoStop}%
\bibitem [{\citenamefont {Fan}\ \emph {et~al.}(2010)\citenamefont {Fan},
  \citenamefont {Wu}, \citenamefont {Bao}, \citenamefont {Bao}, \citenamefont
  {Bardhan}, \citenamefont {Halas}, \citenamefont {Manoharan}, \citenamefont
  {Nordlander}, \citenamefont {Shvets},\ and\ \citenamefont
  {Capasso}}]{Fan2010}%
  \BibitemOpen
  \bibfield  {author} {\bibinfo {author} {\bibfnamefont {J.~A.}\ \bibnamefont
  {Fan}}, \bibinfo {author} {\bibfnamefont {C.}~\bibnamefont {Wu}}, \bibinfo
  {author} {\bibfnamefont {K.}~\bibnamefont {Bao}}, \bibinfo {author}
  {\bibfnamefont {J.}~\bibnamefont {Bao}}, \bibinfo {author} {\bibfnamefont
  {R.}~\bibnamefont {Bardhan}}, \bibinfo {author} {\bibfnamefont {N.~J.}\
  \bibnamefont {Halas}}, \bibinfo {author} {\bibfnamefont {V.~N.}\ \bibnamefont
  {Manoharan}}, \bibinfo {author} {\bibfnamefont {P.}~\bibnamefont
  {Nordlander}}, \bibinfo {author} {\bibfnamefont {G.}~\bibnamefont {Shvets}},
  \ and\ \bibinfo {author} {\bibfnamefont {F.}~\bibnamefont {Capasso}},\
  }\bibfield  {title} {\enquote {\bibinfo {title} {{Self-assembled plasmonic
  nanoparticle clusters}},}\ }\href@noop {} {\bibfield  {journal} {\bibinfo
  {journal} {Science}\ }\textbf {\bibinfo {volume} {328}},\ \bibinfo {pages}
  {1135--1138} (\bibinfo {year} {2010})}\BibitemShut {NoStop}%
\bibitem [{\citenamefont {Liu}\ \emph {et~al.}(2012)\citenamefont {Liu},
  \citenamefont {Mukherjee}, \citenamefont {Bao}, \citenamefont {Brown},
  \citenamefont {Dorfm{\"{u}}ller}, \citenamefont {Nordlander},\ and\
  \citenamefont {Halas}}]{Liu2012}%
  \BibitemOpen
  \bibfield  {author} {\bibinfo {author} {\bibfnamefont {N.}~\bibnamefont
  {Liu}}, \bibinfo {author} {\bibfnamefont {S.}~\bibnamefont {Mukherjee}},
  \bibinfo {author} {\bibfnamefont {K.}~\bibnamefont {Bao}}, \bibinfo {author}
  {\bibfnamefont {L.~V.}\ \bibnamefont {Brown}}, \bibinfo {author}
  {\bibfnamefont {J.}~\bibnamefont {Dorfm{\"{u}}ller}}, \bibinfo {author}
  {\bibfnamefont {P.}~\bibnamefont {Nordlander}}, \ and\ \bibinfo {author}
  {\bibfnamefont {N.~J.}\ \bibnamefont {Halas}},\ }\bibfield  {title} {\enquote
  {\bibinfo {title} {{Magnetic plasmon formation and propagation in artificial
  aromatic molecules}},}\ }\href@noop {} {\bibfield  {journal} {\bibinfo
  {journal} {Nano Lett.}\ }\textbf {\bibinfo {volume} {12}},\ \bibinfo {pages}
  {364--369} (\bibinfo {year} {2012})}\BibitemShut {NoStop}%
\bibitem [{\citenamefont {Shafiei}\ \emph {et~al.}(2013)\citenamefont
  {Shafiei}, \citenamefont {Monticone}, \citenamefont {Le}, \citenamefont
  {Liu}, \citenamefont {Hartsfield}, \citenamefont {Al{\`{u}}},\ and\
  \citenamefont {Li}}]{Shafiei2013}%
  \BibitemOpen
  \bibfield  {author} {\bibinfo {author} {\bibfnamefont {F.}~\bibnamefont
  {Shafiei}}, \bibinfo {author} {\bibfnamefont {F.}~\bibnamefont {Monticone}},
  \bibinfo {author} {\bibfnamefont {K.~Q.}\ \bibnamefont {Le}}, \bibinfo
  {author} {\bibfnamefont {X.~X.}\ \bibnamefont {Liu}}, \bibinfo {author}
  {\bibfnamefont {T.}~\bibnamefont {Hartsfield}}, \bibinfo {author}
  {\bibfnamefont {A.}~\bibnamefont {Al{\`{u}}}}, \ and\ \bibinfo {author}
  {\bibfnamefont {X.}~\bibnamefont {Li}},\ }\bibfield  {title} {\enquote
  {\bibinfo {title} {{A subwavelength plasmonic metamolecule exhibiting
  magnetic-based optical Fano resonance}},}\ }\href@noop {} {\bibfield
  {journal} {\bibinfo  {journal} {Nat. Nanotechnol.}\ }\textbf {\bibinfo
  {volume} {8}},\ \bibinfo {pages} {95--99} (\bibinfo {year}
  {2013})}\BibitemShut {NoStop}%
\bibitem [{\citenamefont {Atwater}\ and\ \citenamefont
  {Polman}(2010)}]{Atwater2010}%
  \BibitemOpen
  \bibfield  {author} {\bibinfo {author} {\bibfnamefont {H.~A.}\ \bibnamefont
  {Atwater}}\ and\ \bibinfo {author} {\bibfnamefont {A.}~\bibnamefont
  {Polman}},\ }\bibfield  {title} {\enquote {\bibinfo {title} {{Plasmonics for
  improved photovoltaic devices.}}}\ }\href@noop {} {\bibfield  {journal}
  {\bibinfo  {journal} {Nat. Mater.}\ }\textbf {\bibinfo {volume} {9}},\
  \bibinfo {pages} {205--13} (\bibinfo {year} {2010})}\BibitemShut {NoStop}%
\bibitem [{\citenamefont {Zhou}\ \emph {et~al.}(2016)\citenamefont {Zhou},
  \citenamefont {Tan}, \citenamefont {Wang}, \citenamefont {Xu}, \citenamefont
  {Yuan}, \citenamefont {Cai}, \citenamefont {Zhu},\ and\ \citenamefont
  {Zhu}}]{Zhou2016}%
  \BibitemOpen
  \bibfield  {author} {\bibinfo {author} {\bibfnamefont {L.}~\bibnamefont
  {Zhou}}, \bibinfo {author} {\bibfnamefont {Y.}~\bibnamefont {Tan}}, \bibinfo
  {author} {\bibfnamefont {J.}~\bibnamefont {Wang}}, \bibinfo {author}
  {\bibfnamefont {W.}~\bibnamefont {Xu}}, \bibinfo {author} {\bibfnamefont
  {Y.}~\bibnamefont {Yuan}}, \bibinfo {author} {\bibfnamefont {W.}~\bibnamefont
  {Cai}}, \bibinfo {author} {\bibfnamefont {S.}~\bibnamefont {Zhu}}, \ and\
  \bibinfo {author} {\bibfnamefont {J.}~\bibnamefont {Zhu}},\ }\bibfield
  {title} {\enquote {\bibinfo {title} {{3D self-assembly of aluminium
  nanoparticles for plasmon-enhanced solar desalination}},}\ }\href@noop {}
  {\bibfield  {journal} {\bibinfo  {journal} {Nat. Photonics}\ }\textbf
  {\bibinfo {volume} {10}},\ \bibinfo {pages} {393--99} (\bibinfo {year}
  {2016})}\BibitemShut {NoStop}%
\bibitem [{\citenamefont {Baranov}\ \emph {et~al.}(2017)\citenamefont
  {Baranov}, \citenamefont {Zuev}, \citenamefont {Lepeshov}, \citenamefont
  {Kotov}, \citenamefont {Krasnok}, \citenamefont {Evlyukhin},\ and\
  \citenamefont {Chichkov}}]{Baranov2017}%
  \BibitemOpen
  \bibfield  {author} {\bibinfo {author} {\bibfnamefont {D.~G.}\ \bibnamefont
  {Baranov}}, \bibinfo {author} {\bibfnamefont {D.~A.}\ \bibnamefont {Zuev}},
  \bibinfo {author} {\bibfnamefont {S.~I.}\ \bibnamefont {Lepeshov}}, \bibinfo
  {author} {\bibfnamefont {O.~V.}\ \bibnamefont {Kotov}}, \bibinfo {author}
  {\bibfnamefont {A.~E.}\ \bibnamefont {Krasnok}}, \bibinfo {author}
  {\bibfnamefont {A.~B.}\ \bibnamefont {Evlyukhin}}, \ and\ \bibinfo {author}
  {\bibfnamefont {B.~N.}\ \bibnamefont {Chichkov}},\ }\bibfield  {title}
  {\enquote {\bibinfo {title} {{All-dielectric nanophotonics: the quest for
  better materials and fabrication techniques}},}\ }\href {\doibase
  10.1364/optica.4.000814} {\bibfield  {journal} {\bibinfo  {journal} {Optica}\
  }\textbf {\bibinfo {volume} {4}},\ \bibinfo {pages} {814} (\bibinfo {year}
  {2017})}\BibitemShut {NoStop}%
\bibitem [{\citenamefont {Kruk}\ and\ \citenamefont
  {Kivshar}(2017)}]{Kruk2017}%
  \BibitemOpen
  \bibfield  {author} {\bibinfo {author} {\bibfnamefont {S.}~\bibnamefont
  {Kruk}}\ and\ \bibinfo {author} {\bibfnamefont {Y.}~\bibnamefont {Kivshar}},\
  }\bibfield  {title} {\enquote {\bibinfo {title} {{Functional Meta-Optics and
  Nanophotonics Governed by Mie Resonances}},}\ }\href {\doibase
  10.1021/acsphotonics.7b01038} {\bibfield  {journal} {\bibinfo  {journal} {ACS
  Photonics}\ }\textbf {\bibinfo {volume} {4}},\ \bibinfo {pages} {2638--2649}
  (\bibinfo {year} {2017})}\BibitemShut {NoStop}%
\bibitem [{\citenamefont {Yang}\ \emph {et~al.}(2017)\citenamefont {Yang},
  \citenamefont {Jiang}, \citenamefont {Zhuo}, \citenamefont {Xie},
  \citenamefont {Wang},\ and\ \citenamefont {Lin}}]{Yang2017}%
  \BibitemOpen
  \bibfield  {author} {\bibinfo {author} {\bibfnamefont {Z.~J.}\ \bibnamefont
  {Yang}}, \bibinfo {author} {\bibfnamefont {R.}~\bibnamefont {Jiang}},
  \bibinfo {author} {\bibfnamefont {X.}~\bibnamefont {Zhuo}}, \bibinfo {author}
  {\bibfnamefont {Y.~M.}\ \bibnamefont {Xie}}, \bibinfo {author} {\bibfnamefont
  {J.}~\bibnamefont {Wang}}, \ and\ \bibinfo {author} {\bibfnamefont {H.~Q.}\
  \bibnamefont {Lin}},\ }\bibfield  {title} {\enquote {\bibinfo {title}
  {{Dielectric nanoresonators for light manipulation}},}\ }\href {\doibase
  10.1016/j.physrep.2017.07.006} {\bibfield  {journal} {\bibinfo  {journal}
  {Phys. Rep.}\ }\textbf {\bibinfo {volume} {701}},\ \bibinfo {pages} {1--50}
  (\bibinfo {year} {2017})}\BibitemShut {NoStop}%
\bibitem [{\citenamefont {Tzarouchis}\ and\ \citenamefont
  {Sihvola}(2018)}]{Tzarouchis2018}%
  \BibitemOpen
  \bibfield  {author} {\bibinfo {author} {\bibfnamefont {D.}~\bibnamefont
  {Tzarouchis}}\ and\ \bibinfo {author} {\bibfnamefont {A.}~\bibnamefont
  {Sihvola}},\ }\bibfield  {title} {\enquote {\bibinfo {title} {{Light
  scattering by a dielectric sphere: Perspectives on the Mie resonances}},}\
  }\href@noop {} {\bibfield  {journal} {\bibinfo  {journal} {Appl. Sci.}\
  }\textbf {\bibinfo {volume} {8}},\ \bibinfo {pages} {184} (\bibinfo {year}
  {2018})}\BibitemShut {NoStop}%
\bibitem [{\citenamefont {Barreda}\ \emph {et~al.}(2019)\citenamefont
  {Barreda}, \citenamefont {Saiz}, \citenamefont {Gonz{\'{a}}lez},
  \citenamefont {Moreno},\ and\ \citenamefont {Albella}}]{Barreda2019}%
  \BibitemOpen
  \bibfield  {author} {\bibinfo {author} {\bibfnamefont {{\'{A}}.~I.}\
  \bibnamefont {Barreda}}, \bibinfo {author} {\bibfnamefont {J.~M.}\
  \bibnamefont {Saiz}}, \bibinfo {author} {\bibfnamefont {F.}~\bibnamefont
  {Gonz{\'{a}}lez}}, \bibinfo {author} {\bibfnamefont {F.}~\bibnamefont
  {Moreno}}, \ and\ \bibinfo {author} {\bibfnamefont {P.}~\bibnamefont
  {Albella}},\ }\bibfield  {title} {\enquote {\bibinfo {title} {{Recent
  advances in high refractive index dielectric nanoantennas: Basics and
  applications}},}\ }\href {\doibase 10.1063/1.5087402} {\bibfield  {journal}
  {\bibinfo  {journal} {AIP Adv.}\ }\textbf {\bibinfo {volume} {9}},\ \bibinfo
  {pages} {040701} (\bibinfo {year} {2019})}\BibitemShut {NoStop}%
\bibitem [{\citenamefont {Paniagua-Dom{\'{i}}nguez}\ \emph
  {et~al.}(2019)\citenamefont {Paniagua-Dom{\'{i}}nguez}, \citenamefont
  {Luk'yanchuk}, \citenamefont {Miroshnichenko},\ and\ \citenamefont
  {S{\'{a}}nchez-Gil}}]{Paniagua-Dominguez2019}%
  \BibitemOpen
  \bibfield  {author} {\bibinfo {author} {\bibfnamefont {R.}~\bibnamefont
  {Paniagua-Dom{\'{i}}nguez}}, \bibinfo {author} {\bibfnamefont
  {B.}~\bibnamefont {Luk'yanchuk}}, \bibinfo {author} {\bibfnamefont
  {A.}~\bibnamefont {Miroshnichenko}}, \ and\ \bibinfo {author} {\bibfnamefont
  {J.~A.}\ \bibnamefont {S{\'{a}}nchez-Gil}},\ }\bibfield  {title} {\enquote
  {\bibinfo {title} {{Dielectric nanoresonators and metamaterials}},}\ }\href
  {\doibase 10.1063/1.5129100} {\bibfield  {journal} {\bibinfo  {journal} {J.
  Appl. Phys.}\ }\textbf {\bibinfo {volume} {126}},\ \bibinfo {pages} {150401}
  (\bibinfo {year} {2019})}\BibitemShut {NoStop}%
\bibitem [{\citenamefont {Mie}(1908)}]{Mie1908}%
  \BibitemOpen
  \bibfield  {author} {\bibinfo {author} {\bibfnamefont {G.}~\bibnamefont
  {Mie}},\ }\bibfield  {title} {\enquote {\bibinfo {title} {{Beitr{\"{a}}ge zur
  Optik tr{\"{u}}ber Medien, speziell kolloidaler Metall{\"{o}}sungen}},}\
  }\href {\doibase 10.1002/andp.19083300302} {\bibfield  {journal} {\bibinfo
  {journal} {Ann. Phys. (Leipzig)}\ }\textbf {\bibinfo {volume} {330}},\
  \bibinfo {pages} {377--445} (\bibinfo {year} {1908})}\BibitemShut {NoStop}%
\bibitem [{\citenamefont {Garc{\'{i}}a-C{\'{a}}mara}\ \emph
  {et~al.}(2013)\citenamefont {Garc{\'{i}}a-C{\'{a}}mara}, \citenamefont
  {G{\'{o}}mez-Medina}, \citenamefont {S{\'{a}}enz},\ and\ \citenamefont
  {Sep{\'{u}}lveda}}]{Garcia-Camara2013}%
  \BibitemOpen
  \bibfield  {author} {\bibinfo {author} {\bibfnamefont {B.}~\bibnamefont
  {Garc{\'{i}}a-C{\'{a}}mara}}, \bibinfo {author} {\bibfnamefont
  {R.}~\bibnamefont {G{\'{o}}mez-Medina}}, \bibinfo {author} {\bibfnamefont
  {J.~J.}\ \bibnamefont {S{\'{a}}enz}}, \ and\ \bibinfo {author} {\bibfnamefont
  {B.}~\bibnamefont {Sep{\'{u}}lveda}},\ }\bibfield  {title} {\enquote
  {\bibinfo {title} {{Sensing with magnetic dipolar resonances in semiconductor
  nanospheres}},}\ }\href {\doibase 10.1364/OE.21.023007} {\bibfield  {journal}
  {\bibinfo  {journal} {Opt. Express}\ }\textbf {\bibinfo {volume} {21}},\
  \bibinfo {pages} {23007} (\bibinfo {year} {2013})}\BibitemShut {NoStop}%
\bibitem [{\citenamefont {Barreda}\ \emph {et~al.}(2015)\citenamefont
  {Barreda}, \citenamefont {Sanz},\ and\ \citenamefont
  {Gonz{\'{a}}lez}}]{Barreda2015}%
  \BibitemOpen
  \bibfield  {author} {\bibinfo {author} {\bibfnamefont {{\'{A}}.~I.}\
  \bibnamefont {Barreda}}, \bibinfo {author} {\bibfnamefont {J.~M.}\
  \bibnamefont {Sanz}}, \ and\ \bibinfo {author} {\bibfnamefont
  {F.}~\bibnamefont {Gonz{\'{a}}lez}},\ }\bibfield  {title} {\enquote {\bibinfo
  {title} {{Using linear polarization for sensing and sizing dielectric
  nanoparticles}},}\ }\href {\doibase 10.1364/oe.23.009157} {\bibfield
  {journal} {\bibinfo  {journal} {Opt. Express}\ }\textbf {\bibinfo {volume}
  {23}},\ \bibinfo {pages} {9157} (\bibinfo {year} {2015})}\BibitemShut
  {NoStop}%
\bibitem [{\citenamefont {G{\'{o}}mez-Medina}\ \emph
  {et~al.}(2011)\citenamefont {G{\'{o}}mez-Medina}, \citenamefont
  {Garc{\'{i}}a-C{\'{a}}mara}, \citenamefont {Su{\'{a}}rez-Lacalle},
  \citenamefont {Gonz{\'{a}}lez}, \citenamefont {Moreno}, \citenamefont
  {Nieto-Vesperinas},\ and\ \citenamefont {Sa{\'{e}}nz}}]{Gomez-Medina2011}%
  \BibitemOpen
  \bibfield  {author} {\bibinfo {author} {\bibfnamefont {R.}~\bibnamefont
  {G{\'{o}}mez-Medina}}, \bibinfo {author} {\bibfnamefont {B.}~\bibnamefont
  {Garc{\'{i}}a-C{\'{a}}mara}}, \bibinfo {author} {\bibfnamefont
  {I.}~\bibnamefont {Su{\'{a}}rez-Lacalle}}, \bibinfo {author} {\bibfnamefont
  {F.}~\bibnamefont {Gonz{\'{a}}lez}}, \bibinfo {author} {\bibfnamefont
  {F.}~\bibnamefont {Moreno}}, \bibinfo {author} {\bibfnamefont
  {M.}~\bibnamefont {Nieto-Vesperinas}}, \ and\ \bibinfo {author}
  {\bibfnamefont {J.~J.}\ \bibnamefont {Sa{\'{e}}nz}},\ }\bibfield  {title}
  {\enquote {\bibinfo {title} {{Electric and magnetic dipolar response of
  germanium nanospheres: interference effects, scattering anisotropy, and
  optical forces}},}\ }\href@noop {} {\bibfield  {journal} {\bibinfo  {journal}
  {J. Nanophotonics}\ }\textbf {\bibinfo {volume} {5}},\ \bibinfo {pages}
  {053512} (\bibinfo {year} {2011})}\BibitemShut {NoStop}%
\bibitem [{\citenamefont {Geffrin}\ \emph {et~al.}(2012)\citenamefont
  {Geffrin}, \citenamefont {Garc{\'{i}}a-C{\'{a}}mara}, \citenamefont
  {G{\'{o}}mez-Medina}, \citenamefont {Albella}, \citenamefont
  {Froufe-P{\'{e}}rez}, \citenamefont {Eyraud}, \citenamefont {Litman},
  \citenamefont {Vaillon}, \citenamefont {Gonz{\'{a}}lez}, \citenamefont
  {Nieto-Vesperinas}, \citenamefont {S{\'{a}}enz},\ and\ \citenamefont
  {Moreno}}]{Geffrin2012}%
  \BibitemOpen
  \bibfield  {author} {\bibinfo {author} {\bibfnamefont {J.}~\bibnamefont
  {Geffrin}}, \bibinfo {author} {\bibfnamefont {B.}~\bibnamefont
  {Garc{\'{i}}a-C{\'{a}}mara}}, \bibinfo {author} {\bibfnamefont
  {R.}~\bibnamefont {G{\'{o}}mez-Medina}}, \bibinfo {author} {\bibfnamefont
  {P.}~\bibnamefont {Albella}}, \bibinfo {author} {\bibfnamefont
  {L.}~\bibnamefont {Froufe-P{\'{e}}rez}}, \bibinfo {author} {\bibfnamefont
  {C.}~\bibnamefont {Eyraud}}, \bibinfo {author} {\bibfnamefont
  {A.}~\bibnamefont {Litman}}, \bibinfo {author} {\bibfnamefont
  {R.}~\bibnamefont {Vaillon}}, \bibinfo {author} {\bibfnamefont
  {F.}~\bibnamefont {Gonz{\'{a}}lez}}, \bibinfo {author} {\bibfnamefont
  {M.}~\bibnamefont {Nieto-Vesperinas}}, \bibinfo {author} {\bibfnamefont
  {J.}~\bibnamefont {S{\'{a}}enz}}, \ and\ \bibinfo {author} {\bibfnamefont
  {F.}~\bibnamefont {Moreno}},\ }\bibfield  {title} {\enquote {\bibinfo {title}
  {{Magnetic and electric coherence in forward- and back-scattered
  electromagnetic waves by a single dielectric subwavelength sphere}},}\
  }\href@noop {} {\bibfield  {journal} {\bibinfo  {journal} {Nat. Commun.}\
  }\textbf {\bibinfo {volume} {3}},\ \bibinfo {pages} {1171} (\bibinfo {year}
  {2012})}\BibitemShut {NoStop}%
\bibitem [{\citenamefont {Tribelsky}\ \emph {et~al.}(2015)\citenamefont
  {Tribelsky}, \citenamefont {Geffrin}, \citenamefont {Litman}, \citenamefont
  {Eyraud},\ and\ \citenamefont {Moreno}}]{Tribelsky2015}%
  \BibitemOpen
  \bibfield  {author} {\bibinfo {author} {\bibfnamefont {M.~I.}\ \bibnamefont
  {Tribelsky}}, \bibinfo {author} {\bibfnamefont {J.-M.}\ \bibnamefont
  {Geffrin}}, \bibinfo {author} {\bibfnamefont {A.}~\bibnamefont {Litman}},
  \bibinfo {author} {\bibfnamefont {C.}~\bibnamefont {Eyraud}}, \ and\ \bibinfo
  {author} {\bibfnamefont {F.}~\bibnamefont {Moreno}},\ }\bibfield  {title}
  {\enquote {\bibinfo {title} {{Small Dielectric Spheres with High Refractive
  Index as New Multifunctional Elements for Optical Devices}},}\ }\href@noop {}
  {\bibfield  {journal} {\bibinfo  {journal} {Sci. Rep.}\ }\textbf {\bibinfo
  {volume} {5}},\ \bibinfo {pages} {12288} (\bibinfo {year}
  {2015})}\BibitemShut {NoStop}%
\bibitem [{\citenamefont {Zhang}\ \emph {et~al.}(2015)\citenamefont {Zhang},
  \citenamefont {Jiang}, \citenamefont {Xie}, \citenamefont {Ruan},
  \citenamefont {Yang}, \citenamefont {Wang},\ and\ \citenamefont
  {Lin}}]{Zhang2015}%
  \BibitemOpen
  \bibfield  {author} {\bibinfo {author} {\bibfnamefont {S.}~\bibnamefont
  {Zhang}}, \bibinfo {author} {\bibfnamefont {R.}~\bibnamefont {Jiang}},
  \bibinfo {author} {\bibfnamefont {Y.~M.}\ \bibnamefont {Xie}}, \bibinfo
  {author} {\bibfnamefont {Q.}~\bibnamefont {Ruan}}, \bibinfo {author}
  {\bibfnamefont {B.}~\bibnamefont {Yang}}, \bibinfo {author} {\bibfnamefont
  {J.}~\bibnamefont {Wang}}, \ and\ \bibinfo {author} {\bibfnamefont {H.~Q.}\
  \bibnamefont {Lin}},\ }\bibfield  {title} {\enquote {\bibinfo {title}
  {{Colloidal Moderate-Refractive-Index $\mathrm{Cu_2O}$ Nanospheres as
  Visible-Region Nanoantennas with Electromagnetic Resonance and Directional
  Light-Scattering Properties}},}\ }\href@noop {} {\bibfield  {journal}
  {\bibinfo  {journal} {Adv. Mater.}\ }\textbf {\bibinfo {volume} {27}},\
  \bibinfo {pages} {7432--7439} (\bibinfo {year} {2015})}\BibitemShut {NoStop}%
\bibitem [{\citenamefont {Ullah}\ \emph {et~al.}(2018)\citenamefont {Ullah},
  \citenamefont {Huang}, \citenamefont {Habib},\ and\ \citenamefont
  {Liu}}]{Ullah2018}%
  \BibitemOpen
  \bibfield  {author} {\bibinfo {author} {\bibfnamefont {K.}~\bibnamefont
  {Ullah}}, \bibinfo {author} {\bibfnamefont {L.}~\bibnamefont {Huang}},
  \bibinfo {author} {\bibfnamefont {M.}~\bibnamefont {Habib}}, \ and\ \bibinfo
  {author} {\bibfnamefont {X.}~\bibnamefont {Liu}},\ }\bibfield  {title}
  {\enquote {\bibinfo {title} {{Engineering the optical properties of
  dielectric nanospheres by resonant modes}},}\ }\href {\doibase
  10.1088/1361-6528/aae4d2} {\bibfield  {journal} {\bibinfo  {journal}
  {Nanotechnology}\ }\textbf {\bibinfo {volume} {29}},\ \bibinfo {pages}
  {505204} (\bibinfo {year} {2018})}\BibitemShut {NoStop}%
\bibitem [{\citenamefont {Tribelsky}\ and\ \citenamefont
  {Miroshnichenko}(2016)}]{Tribelsky2016}%
  \BibitemOpen
  \bibfield  {author} {\bibinfo {author} {\bibfnamefont {M.~I.}\ \bibnamefont
  {Tribelsky}}\ and\ \bibinfo {author} {\bibfnamefont {A.~E.}\ \bibnamefont
  {Miroshnichenko}},\ }\bibfield  {title} {\enquote {\bibinfo {title} {{Giant
  in-particle field concentration and Fano resonances at light scattering by
  high-refractive-indexparticles}},}\ }\href {\doibase
  10.1103/PhysRevA.93.053837} {\bibfield  {journal} {\bibinfo  {journal} {Phys.
  Rev. A}\ }\textbf {\bibinfo {volume} {93}},\ \bibinfo {pages} {053837}
  (\bibinfo {year} {2016})}\BibitemShut {NoStop}%
\bibitem [{\citenamefont {Roll}\ and\ \citenamefont
  {Schweiger}(2000)}]{Roll2000}%
  \BibitemOpen
  \bibfield  {author} {\bibinfo {author} {\bibfnamefont {G.}~\bibnamefont
  {Roll}}\ and\ \bibinfo {author} {\bibfnamefont {G.}~\bibnamefont
  {Schweiger}},\ }\bibfield  {title} {\enquote {\bibinfo {title} {{Geometrical
  optics model of Mie resonances}},}\ }\href {\doibase 10.1364/josaa.17.001301}
  {\bibfield  {journal} {\bibinfo  {journal} {J. Opt. Soc. Am. A}\ }\textbf
  {\bibinfo {volume} {17}},\ \bibinfo {pages} {1301} (\bibinfo {year}
  {2000})}\BibitemShut {NoStop}%
\bibitem [{\citenamefont {Tzarouchis}\ \emph {et~al.}(2016)\citenamefont
  {Tzarouchis}, \citenamefont {Yl{\"{a}}-Oijala},\ and\ \citenamefont
  {Sihvola}}]{Tzarouchis2016}%
  \BibitemOpen
  \bibfield  {author} {\bibinfo {author} {\bibfnamefont {D.~C.}\ \bibnamefont
  {Tzarouchis}}, \bibinfo {author} {\bibfnamefont {P.}~\bibnamefont
  {Yl{\"{a}}-Oijala}}, \ and\ \bibinfo {author} {\bibfnamefont
  {A.}~\bibnamefont {Sihvola}},\ }\bibfield  {title} {\enquote {\bibinfo
  {title} {{Unveiling the scattering behavior of small spheres}},}\ }\href@noop
  {} {\bibfield  {journal} {\bibinfo  {journal} {Phys. Rev. B}\ }\textbf
  {\bibinfo {volume} {94}},\ \bibinfo {pages} {140301(R)} (\bibinfo {year}
  {2016})}\BibitemShut {NoStop}%
\bibitem [{\citenamefont {Tzarouchis}\ \emph {et~al.}(2017)\citenamefont
  {Tzarouchis}, \citenamefont {Yl{\"{a}}-Oijala},\ and\ \citenamefont
  {Sihvola}}]{Tzarouchis2017}%
  \BibitemOpen
  \bibfield  {author} {\bibinfo {author} {\bibfnamefont {D.~C.}\ \bibnamefont
  {Tzarouchis}}, \bibinfo {author} {\bibfnamefont {P.}~\bibnamefont
  {Yl{\"{a}}-Oijala}}, \ and\ \bibinfo {author} {\bibfnamefont
  {A.}~\bibnamefont {Sihvola}},\ }\bibfield  {title} {\enquote {\bibinfo
  {title} {{Resonant scattering characteristics of homogeneous dielectric
  sphere}},}\ }\href@noop {} {\bibfield  {journal} {\bibinfo  {journal} {IEEE
  Trans. Antennas Propag.}\ }\textbf {\bibinfo {volume} {65}},\ \bibinfo
  {pages} {3184--3191} (\bibinfo {year} {2017})}\BibitemShut {NoStop}%
\bibitem [{\citenamefont {Bohren}\ and\ \citenamefont
  {Huffman}(1998)}]{BHBook}%
  \BibitemOpen
  \bibfield  {author} {\bibinfo {author} {\bibfnamefont {C.~F.}\ \bibnamefont
  {Bohren}}\ and\ \bibinfo {author} {\bibfnamefont {D.~R.}\ \bibnamefont
  {Huffman}},\ }\href@noop {} {\emph {\bibinfo {title} {Absorption and
  Scattering of Light by Small Particles}}}\ (\bibinfo  {publisher} {John Wiley
  \& Sons},\ \bibinfo {address} {New York},\ \bibinfo {year}
  {1998})\BibitemShut {NoStop}%
\bibitem [{\citenamefont {Ch{\'{y}}lek}(1973)}]{Chylek1973}%
  \BibitemOpen
  \bibfield  {author} {\bibinfo {author} {\bibfnamefont {P.}~\bibnamefont
  {Ch{\'{y}}lek}},\ }\bibfield  {title} {\enquote {\bibinfo {title}
  {{Large-sphere limits of the Mie-scattering functions}},}\ }\href {\doibase
  10.1364/JOSA.63.000699} {\bibfield  {journal} {\bibinfo  {journal} {J. Opt.
  Soc. Am.}\ }\textbf {\bibinfo {volume} {63}},\ \bibinfo {pages} {699}
  (\bibinfo {year} {1973})}\BibitemShut {NoStop}%
\bibitem [{\citenamefont {Probert-Jones}(1984)}]{Probert-Jones1984}%
  \BibitemOpen
  \bibfield  {author} {\bibinfo {author} {\bibfnamefont {J.~R.}\ \bibnamefont
  {Probert-Jones}},\ }\bibfield  {title} {\enquote {\bibinfo {title}
  {{Resonance component of backscattering by large dielectric spheres}},}\
  }\href {\doibase 10.1364/JOSAA.1.000822} {\bibfield  {journal} {\bibinfo
  {journal} {J. Opt. Soc. Am. A}\ }\textbf {\bibinfo {volume} {1}},\ \bibinfo
  {pages} {822} (\bibinfo {year} {1984})}\BibitemShut {NoStop}%
\bibitem [{\citenamefont {Zhang}\ and\ \citenamefont
  {Jin}(1996)}]{SpecfuncBook}%
  \BibitemOpen
  \bibfield  {author} {\bibinfo {author} {\bibfnamefont {S.}~\bibnamefont
  {Zhang}}\ and\ \bibinfo {author} {\bibfnamefont {J.}~\bibnamefont {Jin}},\
  }\href@noop {} {\emph {\bibinfo {title} {Computation of Special Functions}}}\
  (\bibinfo  {publisher} {John Wiley \& Sons},\ \bibinfo {address} {New York},\
  \bibinfo {year} {1996})\BibitemShut {NoStop}%
\bibitem [{\citenamefont {Mishchenko}\ \emph {et~al.}(2006)\citenamefont
  {Mishchenko}, \citenamefont {Travis},\ and\ \citenamefont
  {Lacis}}]{MishchenkoBook}%
  \BibitemOpen
  \bibfield  {author} {\bibinfo {author} {\bibfnamefont {M.~I.}\ \bibnamefont
  {Mishchenko}}, \bibinfo {author} {\bibfnamefont {L.~D.}\ \bibnamefont
  {Travis}}, \ and\ \bibinfo {author} {\bibfnamefont {A.~A.}\ \bibnamefont
  {Lacis}},\ }\href@noop {} {\emph {\bibinfo {title} {Scattering, Absorption
  and Emission of Light by Small Particles}}},\ \bibinfo {edition} {3rd}\ ed.\
  (\bibinfo  {publisher} {Cambridge University Press \& NASA},\ \bibinfo
  {address} {Cambridge},\ \bibinfo {year} {2006})\BibitemShut {NoStop}%
\bibitem [{\citenamefont {Green}(2008)}]{Green2008}%
  \BibitemOpen
  \bibfield  {author} {\bibinfo {author} {\bibfnamefont {M.~A.}\ \bibnamefont
  {Green}},\ }\bibfield  {title} {\enquote {\bibinfo {title} {{Self-consistent
  optical parameters of intrinsic silicon at 300 K including temperature
  coefficients}},}\ }\href {\doibase 10.1016/j.solmat.2008.06.009} {\bibfield
  {journal} {\bibinfo  {journal} {Sol. Energ. Mat. Sol. Cells}\ }\textbf
  {\bibinfo {volume} {92}},\ \bibinfo {pages} {1305--1310} (\bibinfo {year}
  {2008})}\BibitemShut {NoStop}%
\bibitem [{\citenamefont {Baker(Jr.)}\ and\ \citenamefont
  {Graves-Morris}(1996)}]{PadeBook}%
  \BibitemOpen
  \bibfield  {author} {\bibinfo {author} {\bibfnamefont {G.~A.}\ \bibnamefont
  {Baker(Jr.)}}\ and\ \bibinfo {author} {\bibfnamefont {P.}~\bibnamefont
  {Graves-Morris}},\ }\href@noop {} {\emph {\bibinfo {title} {Pad\'e
  Approximants}}},\ \bibinfo {edition} {2nd}\ ed.,\ \bibinfo {series}
  {Encyclopedia of Mathematics and Its Applications}, Vol.~\bibinfo {volume}
  {59}\ (\bibinfo  {publisher} {Cambridge University Press},\ \bibinfo
  {address} {Cambridge},\ \bibinfo {year} {1996})\BibitemShut {NoStop}%
\bibitem [{\citenamefont {Press}\ \emph {et~al.}(1997)\citenamefont {Press},
  \citenamefont {Teukolsky}, \citenamefont {Vetterling},\ and\ \citenamefont
  {Flannery}}]{NRBook}%
  \BibitemOpen
  \bibfield  {author} {\bibinfo {author} {\bibfnamefont {W.~H.}\ \bibnamefont
  {Press}}, \bibinfo {author} {\bibfnamefont {S.~A.}\ \bibnamefont
  {Teukolsky}}, \bibinfo {author} {\bibfnamefont {W.~T.}\ \bibnamefont
  {Vetterling}}, \ and\ \bibinfo {author} {\bibfnamefont {B.~P.}\ \bibnamefont
  {Flannery}},\ }\href@noop {} {\emph {\bibinfo {title} {Numerical Recipes in
  Fortran 77: The Art of Scientific Computing}}},\ \bibinfo {edition} {2nd}\
  ed.\ (\bibinfo  {publisher} {Cambridge University Press},\ \bibinfo {address}
  {Cambridge},\ \bibinfo {year} {1997})\BibitemShut {NoStop}%
\bibitem [{\citenamefont {Tribelsky}\ \emph {et~al.}(2012)\citenamefont
  {Tribelsky}, \citenamefont {Miroshnichenko},\ and\ \citenamefont
  {Kivshar}}]{Tribelsky2012}%
  \BibitemOpen
  \bibfield  {author} {\bibinfo {author} {\bibfnamefont {M.~I.}\ \bibnamefont
  {Tribelsky}}, \bibinfo {author} {\bibfnamefont {A.~E.}\ \bibnamefont
  {Miroshnichenko}}, \ and\ \bibinfo {author} {\bibfnamefont {Y.~S.}\
  \bibnamefont {Kivshar}},\ }\bibfield  {title} {\enquote {\bibinfo {title}
  {{Unconventional Fano resonances in light scattering by small particles}},}\
  }\href {\doibase 10.1209/0295-5075/97/44005} {\bibfield  {journal} {\bibinfo
  {journal} {Europhys. Lett.)}\ }\textbf {\bibinfo {volume} {97}},\ \bibinfo
  {pages} {44005} (\bibinfo {year} {2012})}\BibitemShut {NoStop}%
\bibitem [{\citenamefont {Arruda}\ \emph {et~al.}(2015)\citenamefont {Arruda},
  \citenamefont {Martinez},\ and\ \citenamefont {Pinheiro}}]{Arruda2015}%
  \BibitemOpen
  \bibfield  {author} {\bibinfo {author} {\bibfnamefont {T.~J.}\ \bibnamefont
  {Arruda}}, \bibinfo {author} {\bibfnamefont {A.~S.}\ \bibnamefont
  {Martinez}}, \ and\ \bibinfo {author} {\bibfnamefont {F.~A.}\ \bibnamefont
  {Pinheiro}},\ }\bibfield  {title} {\enquote {\bibinfo {title} {{Tunable
  multiple Fano resonances in magnetic single-layered core-shell particles}},}\
  }\href {\doibase 10.1103/PhysRevA.92.023835} {\bibfield  {journal} {\bibinfo
  {journal} {Phys. Rev. A}\ }\textbf {\bibinfo {volume} {92}},\ \bibinfo
  {pages} {023835} (\bibinfo {year} {2015})},\ \Eprint
  {http://arxiv.org/abs/1508.02255} {arXiv:1508.02255} \BibitemShut {NoStop}%
\bibitem [{\citenamefont {Miroshnichenko}\ \emph {et~al.}(2015)\citenamefont
  {Miroshnichenko}, \citenamefont {Evlyukhin}, \citenamefont {Yu},
  \citenamefont {Bakker}, \citenamefont {Chipouline}, \citenamefont
  {Kuznetsov}, \citenamefont {Luk'yanchuk}, \citenamefont {Chichkov},\ and\
  \citenamefont {Kivshar}}]{Miroshnichenko2015}%
  \BibitemOpen
  \bibfield  {author} {\bibinfo {author} {\bibfnamefont {A.~E.}\ \bibnamefont
  {Miroshnichenko}}, \bibinfo {author} {\bibfnamefont {A.~B.}\ \bibnamefont
  {Evlyukhin}}, \bibinfo {author} {\bibfnamefont {Y.~F.}\ \bibnamefont {Yu}},
  \bibinfo {author} {\bibfnamefont {R.~M.}\ \bibnamefont {Bakker}}, \bibinfo
  {author} {\bibfnamefont {A.}~\bibnamefont {Chipouline}}, \bibinfo {author}
  {\bibfnamefont {A.~I.}\ \bibnamefont {Kuznetsov}}, \bibinfo {author}
  {\bibfnamefont {B.}~\bibnamefont {Luk'yanchuk}}, \bibinfo {author}
  {\bibfnamefont {B.~N.}\ \bibnamefont {Chichkov}}, \ and\ \bibinfo {author}
  {\bibfnamefont {Y.~S.}\ \bibnamefont {Kivshar}},\ }\bibfield  {title}
  {\enquote {\bibinfo {title} {{Nonradiating anapole modes in dielectric
  nanoparticles}},}\ }\href {\doibase 10.1038/ncomms9069} {\bibfield  {journal}
  {\bibinfo  {journal} {Nat. Commun.}\ }\textbf {\bibinfo {volume} {6}},\
  \bibinfo {pages} {8069} (\bibinfo {year} {2015})}\BibitemShut {NoStop}%
\bibitem [{\citenamefont {Kanwal}(1983)}]{KanwalBook}%
  \BibitemOpen
  \bibfield  {author} {\bibinfo {author} {\bibfnamefont {R.~P.}\ \bibnamefont
  {Kanwal}},\ }\href@noop {} {\emph {\bibinfo {title} {Generalized Functions:
  Theory and Technique}}},\ \bibinfo {series} {Mathematics in Science and
  Engineering}, Vol.\ \bibinfo {volume} {171}\ (\bibinfo  {publisher} {Academic
  Press},\ \bibinfo {address} {New York},\ \bibinfo {year} {1983})\BibitemShut
  {NoStop}%
\bibitem [{\citenamefont {Levenberg}(1944)}]{Levenberg1944}%
  \BibitemOpen
  \bibfield  {author} {\bibinfo {author} {\bibfnamefont {K.}~\bibnamefont
  {Levenberg}},\ }\bibfield  {title} {\enquote {\bibinfo {title} {{A Method for
  the Solution of Certain Non-linear Problems in Least Squares}},}\ }\href@noop
  {} {\bibfield  {journal} {\bibinfo  {journal} {Quart. Appl. Math.}\ }\textbf
  {\bibinfo {volume} {2}},\ \bibinfo {pages} {164--168} (\bibinfo {year}
  {1944})}\BibitemShut {NoStop}%
\bibitem [{\citenamefont {Marquardt}(1963)}]{Marquardt1963}%
  \BibitemOpen
  \bibfield  {author} {\bibinfo {author} {\bibfnamefont {D.~W.}\ \bibnamefont
  {Marquardt}},\ }\bibfield  {title} {\enquote {\bibinfo {title} {{An Algorithm
  for Least-Squares Estimation of Nonlinear Parameters}},}\ }\href@noop {}
  {\bibfield  {journal} {\bibinfo  {journal} {J. Soc. Ind. Appl. Math.}\
  }\textbf {\bibinfo {volume} {11}},\ \bibinfo {pages} {431--441} (\bibinfo
  {year} {1963})}\BibitemShut {NoStop}%
\bibitem [{\citenamefont {Haidu}\ \emph {et~al.}(2011)\citenamefont {Haidu},
  \citenamefont {Fronk}, \citenamefont {Gordan}, \citenamefont {Scarlat},
  \citenamefont {Salvan},\ and\ \citenamefont {Zahn}}]{Haidu2011}%
  \BibitemOpen
  \bibfield  {author} {\bibinfo {author} {\bibfnamefont {F.}~\bibnamefont
  {Haidu}}, \bibinfo {author} {\bibfnamefont {M.}~\bibnamefont {Fronk}},
  \bibinfo {author} {\bibfnamefont {O.~D.}\ \bibnamefont {Gordan}}, \bibinfo
  {author} {\bibfnamefont {C.}~\bibnamefont {Scarlat}}, \bibinfo {author}
  {\bibfnamefont {G.}~\bibnamefont {Salvan}}, \ and\ \bibinfo {author}
  {\bibfnamefont {D.~R.}\ \bibnamefont {Zahn}},\ }\bibfield  {title} {\enquote
  {\bibinfo {title} {{Dielectric function and magneto-optical Voigt constant of
  $\mathrm{Cu_2O}$: A combined spectroscopic ellipsometry and polar
  magneto-optical Kerr spectroscopy study}},}\ }\href {\doibase
  10.1103/PhysRevB.84.195203} {\bibfield  {journal} {\bibinfo  {journal} {Phys.
  Rev. B}\ }\textbf {\bibinfo {volume} {84}},\ \bibinfo {pages} {195203}
  (\bibinfo {year} {2011})}\BibitemShut {NoStop}%
\bibitem [{\citenamefont {Siefke}\ \emph {et~al.}(2016)\citenamefont {Siefke},
  \citenamefont {Kroker}, \citenamefont {Pfeiffer}, \citenamefont {Puffky},
  \citenamefont {Dietrich}, \citenamefont {Franta}, \citenamefont
  {Ohl{\'{i}}dal}, \citenamefont {Szeghalmi}, \citenamefont {Kley},\ and\
  \citenamefont {T{\"{u}}nnermann}}]{Siefke2016}%
  \BibitemOpen
  \bibfield  {author} {\bibinfo {author} {\bibfnamefont {T.}~\bibnamefont
  {Siefke}}, \bibinfo {author} {\bibfnamefont {S.}~\bibnamefont {Kroker}},
  \bibinfo {author} {\bibfnamefont {K.}~\bibnamefont {Pfeiffer}}, \bibinfo
  {author} {\bibfnamefont {O.}~\bibnamefont {Puffky}}, \bibinfo {author}
  {\bibfnamefont {K.}~\bibnamefont {Dietrich}}, \bibinfo {author}
  {\bibfnamefont {D.}~\bibnamefont {Franta}}, \bibinfo {author} {\bibfnamefont
  {I.}~\bibnamefont {Ohl{\'{i}}dal}}, \bibinfo {author} {\bibfnamefont
  {A.}~\bibnamefont {Szeghalmi}}, \bibinfo {author} {\bibfnamefont {E.~B.}\
  \bibnamefont {Kley}}, \ and\ \bibinfo {author} {\bibfnamefont
  {A.}~\bibnamefont {T{\"{u}}nnermann}},\ }\bibfield  {title} {\enquote
  {\bibinfo {title} {{Materials Pushing the Application Limits of Wire Grid
  Polarizers further into the Deep Ultraviolet Spectral Range}},}\ }\href
  {\doibase 10.1002/adom.201600250} {\bibfield  {journal} {\bibinfo  {journal}
  {Adv. Opt. Mater.}\ }\textbf {\bibinfo {volume} {4}},\ \bibinfo {pages}
  {1780--1786} (\bibinfo {year} {2016})}\BibitemShut {NoStop}%
\bibitem [{\citenamefont {Evlyukhin}\ \emph {et~al.}(2012)\citenamefont
  {Evlyukhin}, \citenamefont {Novikov}, \citenamefont {Zywietz}, \citenamefont
  {Eriksen}, \citenamefont {Reinhardt}, \citenamefont {Bozhevolnyi},\ and\
  \citenamefont {Chichkov}}]{Evlyukhin2012}%
  \BibitemOpen
  \bibfield  {author} {\bibinfo {author} {\bibfnamefont {A.~B.}\ \bibnamefont
  {Evlyukhin}}, \bibinfo {author} {\bibfnamefont {S.~M.}\ \bibnamefont
  {Novikov}}, \bibinfo {author} {\bibfnamefont {U.}~\bibnamefont {Zywietz}},
  \bibinfo {author} {\bibfnamefont {R.~L.}\ \bibnamefont {Eriksen}}, \bibinfo
  {author} {\bibfnamefont {C.}~\bibnamefont {Reinhardt}}, \bibinfo {author}
  {\bibfnamefont {S.~I.}\ \bibnamefont {Bozhevolnyi}}, \ and\ \bibinfo {author}
  {\bibfnamefont {B.~N.}\ \bibnamefont {Chichkov}},\ }\bibfield  {title}
  {\enquote {\bibinfo {title} {{Demonstration of magnetic dipole resonances of
  dielectric nanospheres in the visible region}},}\ }\href {\doibase
  10.1021/nl301594s} {\bibfield  {journal} {\bibinfo  {journal} {Nano Lett.}\
  }\textbf {\bibinfo {volume} {12}},\ \bibinfo {pages} {3749--3755} (\bibinfo
  {year} {2012})}\BibitemShut {NoStop}%
\bibitem [{\citenamefont {Kuznetsov}\ \emph {et~al.}(2012)\citenamefont
  {Kuznetsov}, \citenamefont {Miroshnichenko}, \citenamefont {Fu},
  \citenamefont {Zhang},\ and\ \citenamefont {Luk'yanchuk}}]{Kuznetsov2012}%
  \BibitemOpen
  \bibfield  {author} {\bibinfo {author} {\bibfnamefont {A.~I.}\ \bibnamefont
  {Kuznetsov}}, \bibinfo {author} {\bibfnamefont {A.~E.}\ \bibnamefont
  {Miroshnichenko}}, \bibinfo {author} {\bibfnamefont {Y.~H.}\ \bibnamefont
  {Fu}}, \bibinfo {author} {\bibfnamefont {J.}~\bibnamefont {Zhang}}, \ and\
  \bibinfo {author} {\bibfnamefont {B.}~\bibnamefont {Luk'yanchuk}},\
  }\bibfield  {title} {\enquote {\bibinfo {title} {{Magnetic light}},}\ }\href
  {http://www.nature.com/articles/srep00492} {\bibfield  {journal} {\bibinfo
  {journal} {Sci. Rep.}\ }\textbf {\bibinfo {volume} {2}},\ \bibinfo {pages}
  {492} (\bibinfo {year} {2012})}\BibitemShut {NoStop}%
\bibitem [{\citenamefont {Barreda}\ \emph {et~al.}(2017)\citenamefont
  {Barreda}, \citenamefont {Saleh}, \citenamefont {Litman}, \citenamefont
  {Gonz{\'{a}}lez}, \citenamefont {Geffrin},\ and\ \citenamefont
  {Moreno}}]{Barreda2017}%
  \BibitemOpen
  \bibfield  {author} {\bibinfo {author} {\bibfnamefont {{\'{A}}.~I.}\
  \bibnamefont {Barreda}}, \bibinfo {author} {\bibfnamefont {H.}~\bibnamefont
  {Saleh}}, \bibinfo {author} {\bibfnamefont {A.}~\bibnamefont {Litman}},
  \bibinfo {author} {\bibfnamefont {F.}~\bibnamefont {Gonz{\'{a}}lez}},
  \bibinfo {author} {\bibfnamefont {J.~M.}\ \bibnamefont {Geffrin}}, \ and\
  \bibinfo {author} {\bibfnamefont {F.}~\bibnamefont {Moreno}},\ }\bibfield
  {title} {\enquote {\bibinfo {title} {{Electromagnetic polarization-controlled
  perfect switching effect with high-refractive-index dimers and the
  beam-splitter configuration}},}\ }\href@noop {} {\bibfield  {journal}
  {\bibinfo  {journal} {Nat. Commun.}\ }\textbf {\bibinfo {volume} {8}},\
  \bibinfo {pages} {1--8} (\bibinfo {year} {2017})}\BibitemShut {NoStop}%
\bibitem [{\citenamefont {Susman}\ \emph {et~al.}(2017)\citenamefont {Susman},
  \citenamefont {Vaskevich},\ and\ \citenamefont {Rubinstein}}]{Susman2017}%
  \BibitemOpen
  \bibfield  {author} {\bibinfo {author} {\bibfnamefont {M.~D.}\ \bibnamefont
  {Susman}}, \bibinfo {author} {\bibfnamefont {A.}~\bibnamefont {Vaskevich}}, \
  and\ \bibinfo {author} {\bibfnamefont {I.}~\bibnamefont {Rubinstein}},\
  }\bibfield  {title} {\enquote {\bibinfo {title} {{Refractive Index Sensing
  Using Visible Electromagnetic Resonances of Supported $\mathrm{Cu_2O}$
  Particles}},}\ }\href@noop {} {\bibfield  {journal} {\bibinfo  {journal} {ACS
  Appl. Mater. Interfaces}\ }\textbf {\bibinfo {volume} {9}},\ \bibinfo {pages}
  {8177--8186} (\bibinfo {year} {2017})}\BibitemShut {NoStop}%
\bibitem [{\citenamefont {Yavas}\ \emph {et~al.}(2019)\citenamefont {Yavas},
  \citenamefont {Svedendahl},\ and\ \citenamefont {Quidant}}]{Yavas2019}%
  \BibitemOpen
  \bibfield  {author} {\bibinfo {author} {\bibfnamefont {O.}~\bibnamefont
  {Yavas}}, \bibinfo {author} {\bibfnamefont {M.}~\bibnamefont {Svedendahl}}, \
  and\ \bibinfo {author} {\bibfnamefont {R.}~\bibnamefont {Quidant}},\
  }\bibfield  {title} {\enquote {\bibinfo {title} {{Unravelling the Role of
  Electric and Magnetic Dipoles in Biosensing with Si Nanoresonators}},}\
  }\href {\doibase 10.1021/acsnano.9b00572} {\bibfield  {journal} {\bibinfo
  {journal} {ACS Nano}\ }\textbf {\bibinfo {volume} {13}},\ \bibinfo {pages}
  {4582--4588} (\bibinfo {year} {2019})}\BibitemShut {NoStop}%
\bibitem [{\citenamefont {Zhuo}\ \emph {et~al.}(2019)\citenamefont {Zhuo},
  \citenamefont {Cheng}, \citenamefont {Guo}, \citenamefont {Jia},
  \citenamefont {Yu},\ and\ \citenamefont {Wang}}]{Zhuo2019}%
  \BibitemOpen
  \bibfield  {author} {\bibinfo {author} {\bibfnamefont {X.}~\bibnamefont
  {Zhuo}}, \bibinfo {author} {\bibfnamefont {X.}~\bibnamefont {Cheng}},
  \bibinfo {author} {\bibfnamefont {Y.}~\bibnamefont {Guo}}, \bibinfo {author}
  {\bibfnamefont {H.}~\bibnamefont {Jia}}, \bibinfo {author} {\bibfnamefont
  {Y.}~\bibnamefont {Yu}}, \ and\ \bibinfo {author} {\bibfnamefont
  {J.}~\bibnamefont {Wang}},\ }\bibfield  {title} {\enquote {\bibinfo {title}
  {{Chemically Synthesized Electromagnetic Metal Oxide Nanoresonators}},}\
  }\href@noop {} {\bibfield  {journal} {\bibinfo  {journal} {Adv. Opt. Mater.}\
  }\textbf {\bibinfo {volume} {7}},\ \bibinfo {pages} {1900396} (\bibinfo
  {year} {2019})}\BibitemShut {NoStop}%
\bibitem [{\citenamefont {Luk'yanchuk}\ \emph {et~al.}(2017)\citenamefont
  {Luk'yanchuk}, \citenamefont {Paniagua-Dom{\'{i}}nguez}, \citenamefont
  {Minin}, \citenamefont {Minin},\ and\ \citenamefont {Wang}}]{Lukyanchuk2017}%
  \BibitemOpen
  \bibfield  {author} {\bibinfo {author} {\bibfnamefont {B.~S.}\ \bibnamefont
  {Luk'yanchuk}}, \bibinfo {author} {\bibfnamefont {R.}~\bibnamefont
  {Paniagua-Dom{\'{i}}nguez}}, \bibinfo {author} {\bibfnamefont
  {I.}~\bibnamefont {Minin}}, \bibinfo {author} {\bibfnamefont
  {O.}~\bibnamefont {Minin}}, \ and\ \bibinfo {author} {\bibfnamefont
  {Z.}~\bibnamefont {Wang}},\ }\bibfield  {title} {\enquote {\bibinfo {title}
  {{Refractive index less than two: photonic nanojets yesterday, today and
  tomorrow [Invited]}},}\ }\href {\doibase 10.1364/OME.7.001820} {\bibfield
  {journal} {\bibinfo  {journal} {Opt. Mater. Express}\ }\textbf {\bibinfo
  {volume} {7}},\ \bibinfo {pages} {1820} (\bibinfo {year} {2017})}\BibitemShut
  {NoStop}%
\bibitem [{\citenamefont {{Deepak Kallepalli}}\ \emph
  {et~al.}(2013)\citenamefont {{Deepak Kallepalli}}, \citenamefont {Grojo},
  \citenamefont {Charmasson}, \citenamefont {Delaporte}, \citenamefont
  {Ut{\'{e}}za}, \citenamefont {Merlen}, \citenamefont {Sangar},\ and\
  \citenamefont {Torchio}}]{DeepakKallepalli2013}%
  \BibitemOpen
  \bibfield  {author} {\bibinfo {author} {\bibfnamefont {L.~N.}\ \bibnamefont
  {{Deepak Kallepalli}}}, \bibinfo {author} {\bibfnamefont {D.}~\bibnamefont
  {Grojo}}, \bibinfo {author} {\bibfnamefont {L.}~\bibnamefont {Charmasson}},
  \bibinfo {author} {\bibfnamefont {P.}~\bibnamefont {Delaporte}}, \bibinfo
  {author} {\bibfnamefont {O.}~\bibnamefont {Ut{\'{e}}za}}, \bibinfo {author}
  {\bibfnamefont {A.}~\bibnamefont {Merlen}}, \bibinfo {author} {\bibfnamefont
  {A.}~\bibnamefont {Sangar}}, \ and\ \bibinfo {author} {\bibfnamefont
  {P.}~\bibnamefont {Torchio}},\ }\bibfield  {title} {\enquote {\bibinfo
  {title} {{Long range nanostructuring of silicon surfaces by photonic nanojets
  from microsphere Langmuir films}},}\ }\href {\doibase
  10.1088/0022-3727/46/14/145102} {\bibfield  {journal} {\bibinfo  {journal}
  {J. Phys. D: Appl. Phys.}\ }\textbf {\bibinfo {volume} {46}},\ \bibinfo
  {pages} {145102} (\bibinfo {year} {2013})}\BibitemShut {NoStop}%
\end{thebibliography}%

\end{document}